\documentclass[12pt]{article}
\usepackage{bm}
\usepackage{graphicx}
\usepackage{color}
\usepackage{subfigure}
\usepackage{hyperref}
\usepackage{latexsym}
\usepackage{amsthm}
\usepackage{amssymb}
\usepackage{braket}
\usepackage{cite}
\usepackage[rgb]{xcolor}
\usepackage{hyperref}

\usepackage{bbold}
\usepackage{bbm}

\def\be{\begin{equation}}
\def\ee{\end{equation}}
\newcommand{\avg}[1]{\langle{#1}\rangle}

\def\bc{\begin{center}} 
\def\ec{\end{center}}
\def\bea{\begin{eqnarray}}
\def\eea{\end{eqnarray}}

\def\be{\begin{equation}}
\def\ee{\end{equation}}

\def\bc{\begin{center}}
\def\ec{\end{center}}
\def\bea{\begin{eqnarray}}
\def\eea{\end{eqnarray}}

\begin{document}
\title{\bf \Large Information theory\\ of spatial network ensembles}
\author{
Ginestra Bianconi\footnote{ORCID:0000-0002-3380-887X}
\\
\normalsize{The Alan Turing Institute}\\
\normalsize{School of Mathematical Sciences, Queen Mary University of London}\\
}
\date{}

\maketitle
\begin{abstract}{This chapter provides a comprehensive and self-contained discussion of the most recent developments of information theory of networks. Maximum entropy models of networks are the least biased ensembles enforcing a set of constraints and are used in a number of application to produce null model of networks.
 Here maximum entropy ensembles of networks are introduced by distinguishing between microcanonical and canonical ensembles revealing the the non-equivalence of these two classes of ensembles in the case in which an extensive number of constraints is imposed. 
It is very common that network data includes also meta-data describing feature of the nodes such as their position in a real or in an abstract space.
The features of the nodes  can be treated as  latent variables that  determine the cost associated to each link. Maximum entropy network ensembles with latent variables include spatial networks and their generalisation. In this chapter  we cover the case  of transportation networks including airport and rail networks.
Maximum entropy network ensemble satisfy a given set of constraints. However traditional   approaches do not provide any insight on the origin of such constraints. We use  information theory principles to  find the optimal distribution of latent variables in the framework of the classical information theory of networks. This theory  explains the ``blessing of non-uniformity" of data  guaranteeing the efficiency of inference algorithms. 
|{\bf Keywords:} information theory, networks, maximum entropy ensembles, latent variables, spatial networks}
\end{abstract}

\newpage


\section{Introduction}
A complex system is a system that displays emergent properties and  cannot be reduced to the sum of its elements taken in isolation. The most beautiful examples of complex systems are provided in nature by biology and include the  living cell or the brain. Indeed a living cell is able to replicate while this emergent phenomenon cannot be explained by just considering its molecules taken in isolation. Similarly, cognition is function of the brain that cannot be explained considering single neurons taken in isolation.The minimal requirement for having a complex system is therefore that its elements interact in a non-trivial way.
 
Network science \cite{laszlo} is based on the simple yet powerful assumption  that a network capturing all the interaction of a complex system  is   a source of information for the complex system that it represent. In some sense this can be interpreted as the first important step that can allow us to abandon a   reductionist approach and embrace a more comprehensive complex system approach. This approach has demonstrated to be  a fundamental tool  to investigate  the complexity of the vast majority of interacting systems, ranging from transportation networks \cite{barthelemy} to brain networks \cite{bullmore}.

From this point of view the network itself, i.e. the set of nodes and links indicating the complex system interactions, is assumed to carry information about the system function, such as its response to perturbations, and its dynamical behaviour. Indeed is has been widely shown that network topology affects the network dynamical behavior \cite{dorogovtsev2008critical}.
A fully developed information theory of networks would be able to fully unveil this interplay between topology and dynamics and  can be argued to be  among the most promising approaches to understand networks and the complex systems they encode.

Currently maximum entropy network ensembles are considered the pillars of the information theory of networks. Maximum entropy ensembles of networks generate the least biased ensemble given a set of constraints.
They  can be used to generate null models of networks ranging from random graphs \cite{erdds1959random,bollobas} to  null models of networks with  an heterogeneous degree distributions \cite{park2004statistical,molloy1995critical,cimini2019statistical,bianconi2007entropy,bianconi2009,kim2012constructing,del2010efficient}.

Recently it has been shown that maximum entropy ensembles allow for a formulation of an information theory of network that, like statistical mechanics, makes a distinction between canonical and microcanonical network ensembles, i.e. maximum entropy network ensembles  that satisfy soft and hard constraints respectively \cite{anand2009,anand2010,bianconi2018multilayer}.
However, maximum entropy network ensembles of complex networks display a property that is not typically found  in statistical mechanics, i.e. in the relevant case in which  a large (extensive)  number of constraints is imposed, (such as the degree sequence) the canonical  and the microcanonical network ensembles are no longer equivalent \cite{anand2009,anand2010,bianconi2008entropies}. This important theoretical results has also relevant practical effects as for example it reveals differences in the  dynamical processes defined on conjugated canonical and microcanonical network ensembles.

Since many spatial \cite{bianconi2009,Halu,krioukov2010hyperbolic} a non spatial network datasets \cite{caldarelli2002scale,garlaschelli2008maximum} are also enriched by meta-data such as position of nodes in a real or abstract space. This meta-data can be assumed to be connected to the establishment of connections in the network. Since sometimes this meta-data is known, and sometimes it is missing/unknown we refer to this meta-data as latent variables. Given the relevance of these latent variables, a comprehensive  information theory of networks must not only focus on network topology but it should also treat the interplay between network topology and latent variables. This often entails revealing the  interplay between the topology of the network and the underlying geometry of the latent variables.
The important combination of different sources of information (the network topology and the latent variables) can be also turned into a framework for embedding the network into an latent space. Network embedding \cite{Embedding,garcia2019mercator}  is an important tool to extract information from networks that  can be used to devise greedy search algorithms 
\cite{Kleinberg,boguna_navigability} or to predict new drugs in network medicine \cite{menche2015uncovering}.   

Very recently a classical  information theory of spatial and non-spatial networks as been proposed \cite{radicchi2004defining}.
 This approach is based on the formulation of classical network ensembles that reconcile maximum entropy models of networks with maximum entropy models for urban mobility 
~\cite{WILSON1967253,wilson1969use,wilson2011entropy,batty3}.
Most importantly this novel classical information theory approach to networks can be used to define {\em optimal latent variables  distributions}.
While the standard information theory of networks generates maximum entropy networks starting from an imposed set of constraints but does not discuss where these constraints come from, the recently proposed approach can find the optimal distribution of latent variables determining the constraints. This optimal latent variable distribution   guarantees an efficient reconstruction of the network also when the information about the network is compressed by a lossy channel.

Interestingly the classical information theory of networks show that power-law degree distributions \cite{barabasi1999emergence} and  non-uniform distributions of nodes in space are optimal latent variable distributions. These results provide an information theory explanation of the ``blessing of non-uniformity" of data \cite{domingos2012few}.

 In this chapter we present a comprehensive view of information theory of networks including its most relevant recent developments which is complementary with the previous chapters using entropy measures for spatial data \cite{batty1,batty2,batty3}. In Sec. II we  cover the general framework of maximum entropy network ensembles covering canonical and microcanonical ensembles and their non equivalence\cite{anand2009,anand2010}.  In Sec. III we cover the information theory of  spatial networks or in general networks with latent  variables\cite{PNAS, bianconi2009,krioukov2010hyperbolic}. 
Sec. III  is entirely  devoted to the classical information theory of spatial and non-spatial networks \cite{radicchi2004defining}.

We note here that due to space constraints we do not address here the discussion of stochastic block models 
\cite{airoldi2008mixed,peixoto2019bayesian}  an information theory of non-equilibrium networks  \cite{barabasi1999emergence,bianconi2001bose,zhao2011entropy,papadopoulos2012popularity}. 
For the same reasons we do not cover here the very recent   development of a new information theory  of multilayer networks \cite{bianconi_multiplex,bianconi2018multilayer,Menichetti,Halu}, higher-order networks \cite{Owen}  and temporal networks \cite{zhao2011entropy}  although the treatment of these generalised network structures using  information theory  leads to very interesting and non-trivial effects.

\section{Network ensembles}
\subsection{The general framework}
In this section we provide the general framework for canonical  and microcanonical maximum entropy network ensembles. These ensembles are based on the assumption that all nodes are labelled and that we can attribute a probability $P(G)$ to each   network of the ensemble, where the  network $G$ is uniquely identified by its adjacency matrix $\bf a$.
Canonical and microcanonical network ensembles maximize the entropy of the network ensemble given a set of constraints. As such they are the least biased ensemble that meet the requirements imposed by the constraints. The constraints fix the value of network observables  such at the total number of links or the degree sequence, and they can be distinguished in two classes: soft and hard constraints. Soft constraints impose that the observables take  given values in expectation among all networks in the ensemble. Hard constraints impose that the observables take  given values in all the networks of the ensemble. For instance an example of  soft constraint is the constraint that imposes a value for the expected total number of links in the ensemble. The corresponding hard constraint imposes instead that each network of the ensemble has a given total number of links.
Canonical ensembles are maximum entropy network ensembles that satisfy the soft constraints, microcanonical ensembles are maximum entropy network ensembles that satisfy the hard constraints.
Interestingly the canonical and the microcanonical ensembles are not equivalent in the large network limit when the number of imposed constraints is extensive, i.e. it is of the same order of magnitude of the network size.
\subsection{Network  ensembles and their entropy}

        Let us consider the  set $\Omega_G$ of all the  simple networks $G=(V,E)$ formed by the set $V$ of nodes and the set $E$ of links with total number of nodes given by $N$, i.e. $|V|=N$.  
        Let us  assume that the nodes are labelled and each network $G$ is fully described by its adjacency matrix ${\bf a}$. For simplicity we restrict ourself to the case of {\em simple networks}, i.e. networks which are undirected and unweighted and in which the nodes cannot be linked to themselves, i.e. there are no tadpoles.
 A network ensemble assigns to each network $G\in \Omega_G$ a   probability $P(G)$.
Network ensembles can be used to generate null models or for extracting relevant information from a given network dataset  solving inference problems. Network scientists are also interested in measuring the average of different observables  over the network ensemble. Among the possible observables a pivotal role is played by the observation of a given link between the generic nodes $i$ and $j$.
By indicating with $a_{ij}$ the element of the adjacency matrix of the network with $a_{ij}=1$ if $(i,j)\in E$ and $a_{ij}=0$ otherwise, the {\em marginal} $p_{ij}$ is defined as
        \bea
        p_{ij}=\sum_{G\in \Omega_G}P(G)a_{ij}.
        \eea
Therefore the marginal $p_{ij}$ indicates the probability that node $i$ is connected to node $j$ in the ensemble.           
           
          The {\em entropy} $S$ of a network  ensemble, is the Shannon entropy associated to the probability distribution $P(G)$, i.e.
                  \bea
          S=-\sum_{G\in \Omega_G}P(G)\ln P(G),
          \label{eq:1:S}
          \eea
          where if $P(G)=0$ we consider the convention $0\ln 0=0$.
           According to information theory \cite{mackay2003information} the entropy indicates  the logarithm of the number of typical networks in the ensemble. 
        As such the entropy of a network ensemble is a central quantity to evaluate the   information content of the ensemble and can be used to infer important structural properties of the network \cite{PNAS,iacovacci2015mesoscopic}.
         
         According to the maximum entropy principle \cite{cover1999elements} the least biased ensemble of networks satisfying a given set of constraints, i.e. displaying a given set of properties such as number of links, degree sequence ect. is the {\em maximum entropy network ensemble}.

       We distinguish between {\it soft constraints} and {\it hard constraints}. The soft constraints are the constraints satisfied in average over the ensemble of networks. The hard constraints are the constraints satisfied by each network in the ensemble.
     Hard constraints are constraints of the type: 
       \bea
       F_{\mu}(G)=C_{\mu}.
       \label{hard}
       \eea
       A hard constraint can fix, for example, the total number of links $L$, 
       \bea
       F(G)=\sum_{i<j}a_{ij}=L,
       \eea
       or the degree  $k_i$ of a node $i$,
       \bea
       F'_i(G)=\sum_{j=1}^N a_{ij}=k_i.
       \eea
       Soft constraints are constraints of the type: 
       \bea
       \sum_{G}P(G)F_{\mu}(G)=C_{\mu}.
       \label{soft}
       \eea
       A soft constraint can fix, for example, the expected number of links in the ensemble  to the value $L$,
       \bea
        \sum_{G\in \Omega_G}P(G)\left[\sum_{i<j}a_{ij}\right]=\sum_{i<j}p_{ij}=L,\nonumber \\
       \eea or the expected degree $k_i$ of node $i$, i.e.
      \bea
      \sum_{G\in \Omega_G}P(G)\left[\sum_{j=1}^Na_{ij}\right]=\sum_{j=1}^Np_{ij}=k_i.
        \eea

 The maximum entropy network ensemble that satisfies a set of soft constraints of the type defined in Eq. (\ref{soft}) is called {\em canonical network ensemble}, the maximum entropy network ensemble that satisfies a set of hard contraints of the type defined in Eq. (\ref{hard}) is called {\em micro-canonical network ensemble}. This terminology was coined in Ref. \cite{anand2009}  and is borrowed by statistical mechanics where there is a similar distinction between ensembles of particles kept at   a given total energy $E$ or  kept in a thermal bath fixing the expected energy $\avg{E}$ of the ensemble \cite{Huang}.
 The fact that the statistical mechanics framework has a strictly related counterpart in network science might seem surprising at first but  it is related to the deep relation of the maximum entropy principle with statistical mechanics \cite{jaynes1957information}.
        
        \subsection{Canonical ensemble}
        \label{sec:canonical}
         In this paragraph we provide the general framework for deriving the expression of the probability $P(G)$ of a network $G$ in the canonical ensemble. This framework is very general. It can  be applied to structural constraints only fixing  topological properties of the network, (such a number of links, degree sequence and so on) and can be applied as well to  spatial networks where the   constraints relate the topology to the  underlying geometrical properties of the spatial networks  (see Sec.\ref{sec:spatial}).
         
        The probability $P(G)$ in a canonical network ensemble satisfying $\hat{P}$ soft constraints 
       \bea
         \sum_{G\in \Omega_G}P(G)F_{\mu}(G)=C_{\mu},
         \label{const}
       \eea 
       with $\mu=1,2\ldots, \hat{P}$ is the  {\em Gibbs measure} given by 
       \bea
       P(G)=\frac{1}{Z}e^{-\sum_{\mu=1}^P \lambda_{\mu}F_{\mu}(G)},
       \label{ex}
       \eea
       where $Z$ is the normalization sum and where the Lagrangian multipliers $\lambda_{\mu}$ are fixed by the constraints 
       in Eq. $(\ref{const})$.
         Given the exponential form of the Gibbs measure, the networks in the canonical ensembles are also called {\em exponential random networks}.
         
        In order to derive Eq.(\ref{ex})  we maximize the entropy of the ensemble $S$ (given by Eq.(\ref{eq:1:S})) under the constraints given by Eqs. $(\ref{const})$ and the condition that $P(G)$ is normalized.
        We define a functional ${\cal F}$ in which we have introduced the Lagrangian multipliers $\lambda_{\mu}$, and $\nu$
        \bea
      \hspace{-4mm}  {\cal F}=-\sum_{G\in \Omega_G}P(G)\ln P(G)-\sum_{\mu}\lambda_{\mu} \left[\sum_{G\in \Omega_G}P(G)F_{\mu}(G)-C_{\mu}\right]-\nu \left[\sum_{G\in \Omega_G}P(G)-1\right].\nonumber
        \eea
        Performing the functional derivative,  and putting the functional derivative to zero, 
        \bea
        \frac{\partial {\cal F}}{\partial P(G)}=-\ln P(G)-1-\sum_{\mu}\lambda_{\mu}F_{\mu}(G)-\nu=0,
        \eea
        we obtain
        \bea
        P(G)=e^{-\sum_{\mu}\lambda_{\mu}F_{\mu}(G)-\nu-1}.
        \eea
        Imposing the normalization condition i.e. putting $\partial \mathcal{F}/\partial \nu=0$ we derive the expression of the Gibbs measure
        \bea
        P(G)=\frac{1}{Z}e^{-\sum_{\mu}\lambda_{\mu}F_{\mu}(G)},
        \eea
        where 
        \bea
        Z=e^{\nu+1}=\sum_{G\in \Omega_G}e^{-\sum_{\mu}\lambda_{\mu}F_{\mu}(G)},
        \eea
        and where the Lagrangian multipliers $\lambda_{\mu}$ are fixed imposing the constraints obtained by putting $\partial \mathcal{F}/\partial \lambda_{\mu}=0$, i.e.
        \bea
        \sum_{G}P(G)F_{\mu}(G)=\sum_{G}\frac{1}{Z}e^{-\sum_{\mu}\lambda_{\mu}F_{\mu}(G)}F_{\mu}(G)=C_{\mu},
        \eea
       for $1\leq \mu\leq \hat{P}.$
       
        \subsection{Examples of canonical network ensembles}
        \label{sec:can:ex}
        We are now in the position to consider some specific examples of canonical ensembles with topological constraints. In particular here we will  derive the entropy and the marginals of the random graph $\mathbbm{G}(N,p)$ \cite{erdds1959random,bollobas} and the canonical network ensemble in which we fix the expected degree of each node of the network \cite{park2004statistical}.
        
     The $\mathbbm{G}(N,p)$ is the  canonical network ensemble in which the average number of links are fixed to $L=pN(N-1)/2$.
        Therefore the ensemble satisfies only the single global constraint ($\hat{P}=1$) of the type defined in Eq.$(\ref{const})$, i.e.
        \bea
        \sum_{G\in \Omega_G}P(G)\left[\sum_{i<j}a_{ij}\right]=L.
        \eea
According to the general treatment of the canonical ensemble discussed in the previous paragraph,  the  probability $P(G)$ of a network $G$  in this ensemble is given by the Gibbs measure defined in Eq.(\ref{ex}). Therefore we obtain that in this case $P(G)$ reads 
        \bea
        P(G)=\frac{1}{Z}e^{-\lambda\sum_{i<j}a_{ij}}.
        \label{pL}
        \eea
        The normalization sum $Z$ is given by 
        \bea
        \hspace{-8mm}Z=\sum_{\{a_{ij}\}}e^{-\lambda\sum_{i<j}a_{ij}}=\left(\sum_{a_{ij}=0,1}e^{-\lambda a_{ij}}\right)^{N(N-1)/2}=
        (1+e^{-\lambda})^{N(N-1)/2},
        \eea
       and the marginal probability $p_{ij}$ of each link $(i,j)$ is the same for every link, i.e. $p_{ij}=p$ and given by 
        \bea
        p=\sum_{\{a_{rs}\}}a_{ij}P(G)=\frac{e^{-\lambda}}{1+e^{-\lambda}}.
        \eea
        The quantity $p$ or equivalently the quantity $\lambda$, is fixed by the condition imposed by Eq.(\ref{pL}) with 
        $pN(N-1)/2=L.$
        It follows that  the probability $P(G)$ of a network $G$ in the  $\mathbbm{G}(N,p)$ ensemble,  is given by 
        \bea
        P(G)=p^{L}(1-p)^{N(N-1)/2-L}.
        \eea
        The entropy $S$ of the $\mathbbm{G}(N,p)$ ensemble can be simply written as 
          \bea
        S=-\frac{N(N-1)}{2}\left[p\ln p+\sum_{i<j}(1-p)\ln(1-p)\right].
        \eea
        As a second example of canonical network ensemble we consider the ensemble \cite{park2004statistical} in which we  we fix the expected degree sequence $\{k_i\}_{i=1,2,\ldots, N}$.
        This network ensemble satisfies the $\hat{P}=N$ soft constraints  of the type defined in Eq. (\ref{const}),
        \bea
        \sum_{G\in \Omega_G}P(G)\left[\sum_{j=1}^Na_{ij}\right]=\sum_{j=1}^N p_{ij}=k_i,
        \label{k:cost}
        \eea
        with $i=1,2\ldots, N$.
       The Gibbs measure given by Eq. (\ref{ex})  reads explicitly for this  ensemble reads explicitly
        \bea
        P(G)=\frac{1}{Z}e^{-\sum_i \lambda_i \sum_{j}a_{ij}},
        \eea
        where the normalization constant $Z$ is given by 
        \bea
        Z&=&\sum_{\{a_{ij}\}}e^{-\sum_i \lambda_i \sum_{j}a_{ij}}=\sum_{\{a_{ij}\}}e^{\sum_{i<j}(\lambda_i+\lambda_j) a_{ij}}=\prod_{i<j}\left(1+e^{-\lambda_i-\lambda_j}\right),
        \eea
        where $\lambda_i$ are the Lagrangian multipliers fixing the $\hat{P}=N$ constraints given by Eq.(\ref{k:cost})
        The marginal probability $p_{ij}$ of a link between node $i$ and node $j$ is given by 
        \bea
        p_{ij}&=&\sum_{\{a_{rs}\}}a_{ij}\frac{1}{Z}e^{\sum_{r<s}(\lambda_r+\lambda_s) a_{rs}}=\frac{e^{-\lambda_i-\lambda_j}}{1+e^{-\lambda_i-\lambda_j}}.
        \label{pij}
        \eea
        This marginal has the Fermi-Dirac form \cite{Huang} where the Fermi-Dirac distribution in statistical mechanics characterizes the occupation number of quantum Fermi particles. The Fermi particles are quantum particles that  obey the exclusion principle, i.e. there cannot be more than one particle in each energy state. This has a parallelism in the canonical network ensemble in which no more than one link can connect a given pair of nodes $(i,j)$, i.e. the adjacency matrix has binary elements $a_{ij}\in \{0,1\}$.
        The expression (\ref{pij}) for the marginal probability $p_{ij}$ has also another notable property, meaning that  it does not factorize into contributions dependent of node $i$ and node $j$. This property of the marginal is connected to the fact that this model in general displays degree correlations, i.e. the degree of two linked nodes is not uncorrelated, but there is a tendency of high degree nodes to connect preferentially to low degree ones. These disassortative correlations emerging in this maximum entropy model fixing exclusively the expected degree sequence are called {\em natural correlations} \cite{garlaschelli2008maximum,bianconi2007entropy,bianconi2009}. 
        
By using the expression (\ref{pij}) for the marginals the constraints (\ref{k:cost}) that fix the value of the Lagrangian multipliers $\lambda_i$ can be also written as
        \bea
        \sum_{j=1}^N p_{ij}=\sum_{j=1} \frac{e^{-\lambda_i-\lambda_j}}{1+e^{-\lambda_i-\lambda_j}}=k_i.
        \label{l:cos}
        \eea
        The probability $P(G)$ of  network factorizes in this case in terms depending on single links, i.e.
        \bea
        P(G)=\prod_{i<j}p_{ij}^{a_{ij}}(1-p_{ij})^{1-a_{ij}}.
        \eea
        Consequently the entropy $S$ given by Eq.(\ref{eq:1:S}) takes the form
        \bea
        S=-\sum_{i<j}p_{ij}\ln p_{ij}-\sum_{i<j}(1-p_{ij})\ln(1-p_{ij}).
        \eea
        We have already mentioned that in this ensemble there are in general natural correlations. These correlations are due to the fact that when we try to impose the given expected degree sequence, if we have nodes with very high degree, also called hub nodes, it is more challenging to find network realizations in which the expected degree sequence is preserved but the network remains simple.
        This suggests that if we impose a cutoff to the maximum degree of the network, these correlations might be washed out.
        Let us test this hypothesis by enforcing a so-called {\em structural cutoff}  \cite{boguna2004cut,bianconi2009,bianconi2007entropy}.
        We  say that a network has a structural cutoff if 
         the maximal degree $K$ of the network is below  the structural cutoff $K_s$, i.e.
        if 
        \bea
        K\ll K_s=\sqrt{\avg{k}N}. 
        \label{structural}
        \eea
        Let us   consider the canonical network ensemble with fixed expected degree sequence  where the networks have the structural cutoff, i.e. they obey Eq.(\ref{structural}).
        In this case it can be seen that Eq. (\ref{l:cos}) has   in the first order  approximation solution
        \bea
        e^{-\lambda_i}=\frac{k_i}{\sqrt{\avg{k}N}}\ll 1,
        \eea
        and consequently the marginal probability $p_{ij}$ factorizes, i.e.
        \bea
        p_{ij}=\frac{k_ik_j}{\avg{k}N}.
        \eea 
        This result implies that, while in general  canonical network ensembles enforcing the expected degree sequence display natural correlations, if we restrict our interest to expected degree sequence with structural cutoff then, in first approximation, the resulting networks of the ensemble are uncorrelated. Since uncorrelated networks are fundamental to study dynamical processes on networks \cite{barrat2008dynamical,dorogovtsev2008critical}, this result sheds lights on the best way to generate such networks. 
        \begin{figure}
\centering 
\includegraphics[width=0.95\textwidth]{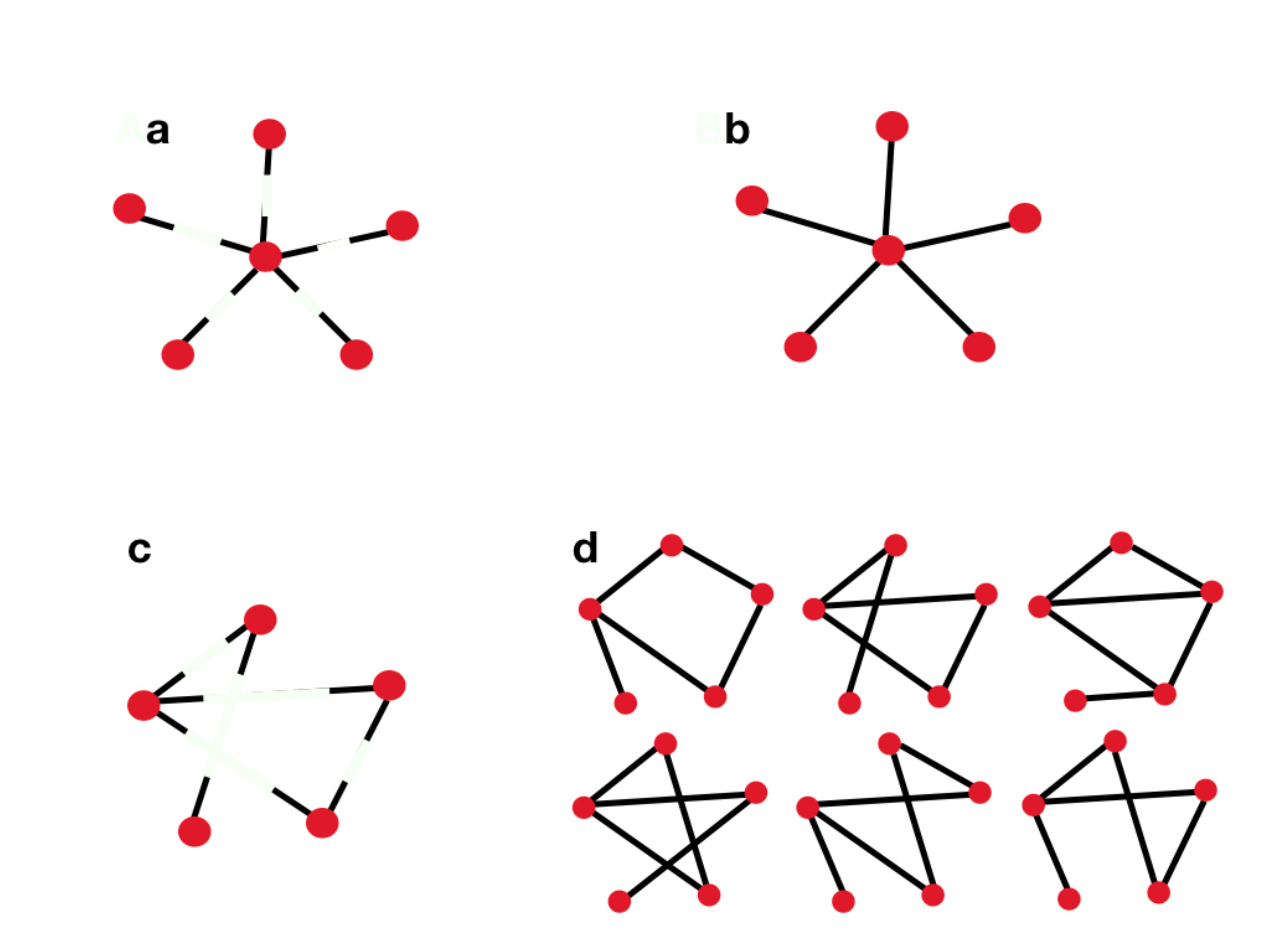}
\caption{The configuration model is generated by starting from a set of $N$ nodes with each node connected to $k_i$ stubs (panels a and c). The  networks in the configuration model are generated by matching the stubs   with the condition of forming a simple network (panels b and d).  This figures shows that two degree sequences corresponding to the same total  number of links  can give rise to microcanonical ensembles of different cardinality   $\mathcal{N}$ and  therefore different entropy $\Sigma$. }
\label{fig:conf}
\end{figure}

          \subsection{Microcanonical ensemble}
          
          Microcanonical network ensembles are maximum entropy networks obeying a set of hard constraints of the type defined in Eq. (\ref{hard}).
      
          Let us indicate by $\Sigma$ the entropy of a microcanonical ensemble   where we use a different notation with respect to the entropy $S$ of  canonical ensemble for later notational convenience. The entropy $\Sigma$ is defined as  in Eq. (\ref{eq:1:S}), i.e.
          \bea
             \Sigma=-\sum_{G\in \Omega_G}P(G)\ln P(G),
          \eea
              By maximizing  $\Sigma$
          given a set of $\hat{P}$ constraints of the type defined in Eq. (\ref{hard}) we obtain that the probability $P(G)$ is uniform over all the networks obeying the mentioned constraints. In other words the probability $P(G)$ of a network $G$ is given by 
                  \bea
        P(G)=\frac{1}{\mathcal{N}}\prod_{\mu=1}^P\hat{\delta}_{F_\mu(G),C_\mu},
        \eea
        where $\hat{\delta}_{r,s}$ indicates the Kornecker delta, i.e. $\hat{\delta}_{r,s}=1$ is $r=s$ and $\hat{\delta}_{r,s}=0$ otherwise.
         Moreover here $\mathcal{N}$ indicates the number of all network $G$ that satisfy the imposed hard constraints, i.e.
        \bea
        \mathcal{N}=\sum_{G\Omega_G} \prod_{\mu=1}^{\hat{P}}\hat{\delta}_{F_\mu(G),C_\mu}.
        \eea
        It follows that for the microcanonical network ensemble the entropy 
        is given by 
        \bea
          \Sigma= \ln \mathcal{N}.
        \eea
        Therefore the entropy $\Sigma$ is given by the logarithm of the cardinality $\mathcal{N}$ of the set of networks that satisfy the hard constraints and as such it is  a fundamental measure of the information content encoded in the constraints. In particular the larger is $\mathcal{N}$, the larger is the entropy $\Sigma$ and the constraints are less informative. If instead the constraints are more difficult to satisfy, i.e. they carry relevant information, then $\mathcal{N}$ and consequently $\Sigma$ take smaller values.
Examples of microcanonical network ensembles are the random graph $\mathbbm{G}(N,L)$ of all networks with $N$ nodes and $L$ links   and the configuration model \cite{molloy1995critical} in which each network of the ensemble obeys the  same degree sequence.
Interestingly we note here that not all hard constraints can be satisfied. For instance  a constraint enforcing a degree sequence whose sum of all degrees is odd cannot be satisfied by any network as the sum of all degrees in a network should be equal to twice the number of links, and therefore must be even.
There are also other less trivial conditions that a degree sequence needs to satisfy in order to be graphical, i.e. in order to be realised at least in a network. 
If one wants to make sure that a degree sequence is graphical one can check the hypothesis of the Erd\"os-Gallai theorem \cite{gallai,del2010efficient,kim2012constructing} which set the necessary and sufficient conditions for graphicality.
Alternatively one can always start from a degree sequence observed in  a real network producing null models using the configuration model.

Let us assume that we have a graphical degree sequence then the networks in the configuration model can be generated by starting from a set of nodes where to each nodes we assign $k_i$ stubs and then randomly matching the stub making sure no tadpoles of multiplex edges are generated in the process (see Figure \ref{fig:conf}).
As it is visible for Figure $\ref{fig:conf}$ different  degree sequences give rise to configuration models with different cardinality $\mathcal{N}$  and consequently different entropy $\Sigma$. In particular the degree sequence compatible with a star network can be matched in a unique way, i.e. has entropy $\Sigma=0$ while degree distribution that are more homogeneous can allow for an entropy $\Sigma>0$ (see Figure \ref{fig:conf}).

\begin{figure}
      \centering 
\includegraphics[width=0.95\textwidth]{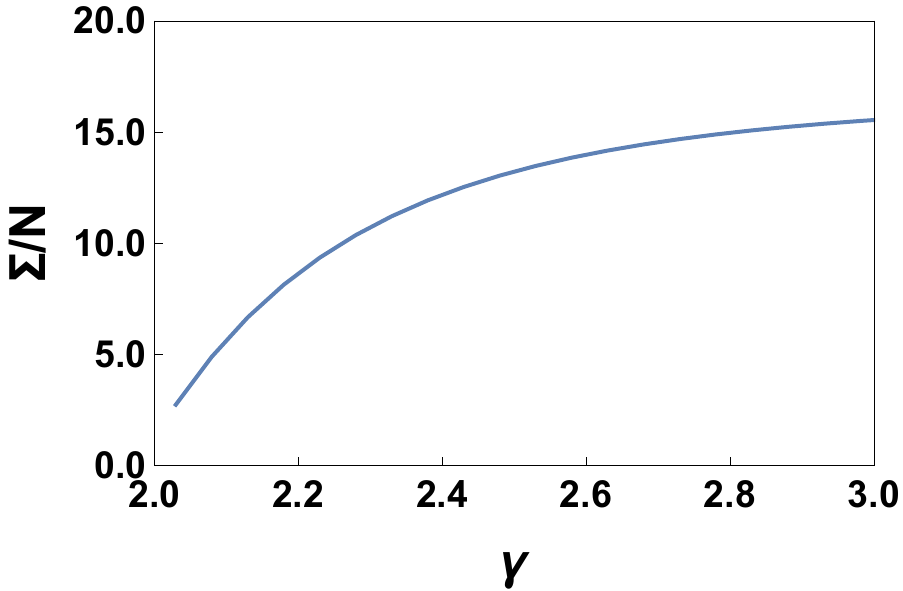}
\caption{The normalized entropy $\Sigma/N$ obtained for the configuration model of uncorrelated scale free networks with degree distribution $P(k)=Ck^{-\gamma}$ is plotted versus the power-law exponent $\gamma>2$. As $\gamma$ decreases  the network is more dominated by hub nodes, leading to a decrease of   the entropy $\Sigma$.}
\label{fig:Bender}
\end{figure}

The problem of graphicality is  not an obstacle in proving asymptotic results for the number of networks with given degree sequence, as these asymptotic results are obtained in the large network limit where we can allow few stubs not to be matched.
For uncorrelated sparse networks, Bender and Canfield proved \cite{Bender} that the asymptotic number of networks in the configuration model is given by 
     \bea
     \mathcal{N}\simeq \frac{L!!}{\prod_{i=1}^Nk_i!}\exp\left[-\frac{1}{4}\left(\frac{\avg{k^2}}{\avg{k}}\right)^2\right].
     \eea
     This result has been also derived using statistical mechanics methods in Refs. \cite{bianconi2008entropies,bianconi2007entropy,bianconi2009,anand2010} 
     The Bender-Canfield formula implies that $\mathcal{N}$ and therefore also $\Sigma$ not only depend on the number of hard constraints imposed but they also depend on the type of constraints imposed.
     In particular for scale-free graphs with constant average degree $\avg{k}$ and degree distribution $P(k)=Ck^{-\gamma}$ the entropy $\Sigma$ is an increasing function of the power-law exponent $\gamma>2$ (see Figure \ref{fig:Bender}).
     This has the intuitive meaning that the broader is the degree distribution, the more dominated the network  is by large degree nodes also called {\em hubs}. Therefore the smaller is the exponent $\gamma$, the broader is the degree distribution and  the more ``star-like"  the network is in proximity to the hubs, leading to  a smaller value of  the entropy $\Sigma$.
     
  \subsection{Non-equivalence of the microcanonical and canonical ensembles}

     The microcanonical and canonical network ensembles enforcing the constraints in Eq.(\ref{hard}) and Eq.(\ref{soft}) with the same set of $F_{\mu}(G)$ and $C_{\mu}$ are called {\em conjugated}.
         If the constraints are subextensive (as the total number of links and the expected total number of links) then the microcanonical and the canonical ensembles are asymptotically equivalent meaning that most of their statistical properties are the same. This is what happens for instance for the $\mathbb{G}(N,p)$ and the $\mathbb{G}(N,L)$ ensembles of random networks.
         However if the number of constraints $\hat{P}$ is  extensive, i.e. they are of the same order as the total number of nodes $N$ of the network,  the canonical and the microcanonical ensembles are no longer equivalent. This happens for instance when we fix the given degree sequence or the expected degree sequence. In this latter case, the relation between the entropy of the canonical network ensemble $S$ and the entropy $\Sigma$ of the conjugated microcanonical network ensemble is given  by \cite{anand2009,anand2010,bianconi2008entropies}
         \bea
         \Sigma=S+N\sum_{k}P(k)\ln \pi_k(k).
         \eea
         where $P(k)$ is the degree distribution of the network and
         \bea
         \pi_k(k)=\frac{1}{k!}k^k e^{-k}.
         \eea
         Since $ \pi_k(k)\leq 1$, the entropy of the microcanonical ensemble $\Sigma$ is smaller than the entropy of the canonical ensemble $S$, i.e.
         \bea
         \Sigma<S,
         \eea
         with $\Sigma-S=\mathcal{O}(N)$. Therefore the difference between the two entropies is not negligible \cite{anand2009,anand2010,bianconi2008entropies}.
        Since  $\Sigma$ and $S$ differs by an extensive (i.e. not negligible) quantity this indicates  the  {\em non-equivalence of the canonical and microcanonical ensembles}.
         One revealing example of this non-equivalence  is the notable difference \cite{anand2009} between a random regular network in which each node has the same exact degree and a Poisson network in which each node, in average has the same degree. The difference between the two ensemble is notable and is also reflected in the different behavior of some dynamical processes and critical phenomena\cite{barrat2008dynamical,dorogovtsev2008critical}.
         For instance the percolation threshold $\hat{p}_c$ of a uncorrelated network with given degree distribution is given by 
         \bea
         \hat{p}_c\frac{\avg{k(k-1)}}{\avg{k}}=1,
         \eea
         where $\avg{\ldots}$ means the average over the degree distribution of the network.
         Therefore for  a regular network in which all the nodes have degree $c$, i.e. $P(k)=\hat{\delta}_{k,c}$ with get 
         \bea
         \hat{p}_c=\frac{1}{c-1}.
         \eea
         However in a Poisson network with average degree $\avg{k}=c$ we have $\avg{k(k-1)}=c^2$ therefore we get a different percolation threshold 
         \bea
         \hat{p}_c=\frac{1}{c}.
         \eea
Note that the non equivalence of the microcanonical and canonical network ensemble is a general property of maximum entropy network ensembles that has been shown to extend also to multiplex networks   \cite{bianconi_multiplex} and simplicial complexes \cite{Owen}.

         \section{Canonical spatial network ensembles}
\label{sec:spatial}
\subsection{Meta-data: spatial coordinates and latent variables}
     
Network data often comes together with meta-data, consisting of a string of variables associated to the nodes and/or links. These variables can be node coordinates on some real embedding space of a spatial network \cite{barthelemy2011spatial}, like in transportation networks, but they can also be variables  that  live in some hidden space. 
Examples of a hidden metric spaces are social spaces in which pair of nodes are associated to a social distance  that  represents and drives the establishment of network interactions \cite{boguna2004models}.
The general notion of hidden metric spaces is fundamental for formulating efficient embedding algorithms where the position of nodes in the embedding space indicates an higher chance that the two nodes are connected or that they belong to the same community.
Embedding algorithms can be used to infer greedy algorithms for network navigability  \cite{boguna_navigability,Embedding,garcia2019mercator} or for  inferring information from network dataset including missing information and one of the most relevant example of this type of approach is constituted by network medicine \cite{menche2015uncovering}.
However in order to have solid statistical approaches for network embedding it is crucial to have a full understanding of the maximum entropy models that assume that the position of nodes in space is known, i.e. spatial networks.
In this section by presenting the theory of spatial networks we will encounter also another class of latent variables associated to nodes. These latent variables are scalar variables for which no obvious definition of metric exist. These might include the GDP of a country in the World Trade Networks \cite{garlaschelli2005structure}, 
assets and liabilities of a bank in financial networks \cite{anand2015filling}
or any other variable that might constitute a proxy for the degree of the nodes \cite{caldarelli2002scale,squartini2018reconstruction}. Despite these latent variables appear to have a different nature with respect to the metric ones, one can consider enriched spatial network ensembles including  a combination variables some associated with a metric some not.
In the well known model for maximum entropy hyperbolic networks \cite{krioukov2010hyperbolic,Serrano_trade}  the non-Euclidean hyperbolic geometry is used to combine  metric variables and non metric variables  in an hyperbolic Poicar\'e disk providing the advantage of a global geometric representation of the interaction pattern in the network.

\subsection{Entropy of  spatial network ensembles}

Consider a simple network of $N$ nodes.
Let us assume that each node   $i$  is assigned a set of  latent variables encoded in the vector  ${\bf x}_i$.
The vector ${\bf x}_i$ might indicate the actual geographical position of nodes in space, a position of nodes in a generalised space (social distance in social networks or any suitable hidden embedding space).

Let us indicate with $\bf x$ the set of  positions ${\bf x}_i$ of every node $i$ of the network,i.e. 
\bea
{\bf x}=\left\{{\bf x}_1,{\bf x}_3,\ldots, {\bf x}_N\right\}.
\label{sp:x}
\eea
The  networks with ensembles with metric latent variables  are spatial networks.However we will see that also non-metric scalar latent variables can be treated using the general approach that we outline here. The only case we will not consider,in this chapter due to space limitation,  is the case of discrete latent variables giving rise to the well-known stochastic-block model \cite{airoldi2008mixed,peixoto2019bayesian}.
 
A spatial network ensemble is an ensemble  where the probability of a given network depends on the positions $\bf x$ of the nodes in space, also called {\em latent variables} given by Eq.(\ref{sp:x}).
Therefore a   spatial ensemble is determined by the probability $P(G|{\bf x})$ associated to each network $G$ between a set of $N$ nodes with latent variables ${\bf x}$.
     The {\em entropy} $S$ of the spatial network ensemble, is the Shannon entropy associated to the probability distribution $P(G|{\bf x})$, i.e.
                  \bea
          S=-\sum_{G}P(G|{\bf x})\ln P(G|{\bf x}).
          \label{sp:S}
          \eea
          
The maximum entropy spatial networks are obtained by maximizing the entropy $S$ of spatial networks given by Eq.(\ref{sp:S}) given a set of $\hat{P}$ constraints.
In this section, for simplicity we will focus exclusively on canonical networks ensembles satisfying as set of constraints of the type 
\bea
\sum_{G\Omega_G}P(G|{\bf x})F_{\mu}(G)=C_{\mu},
\label{sp:const}
\eea
for $1\leq \mu\leq \hat{P}$.
\begin{figure}[h]
\centering 
\includegraphics[width=0.95\textwidth]{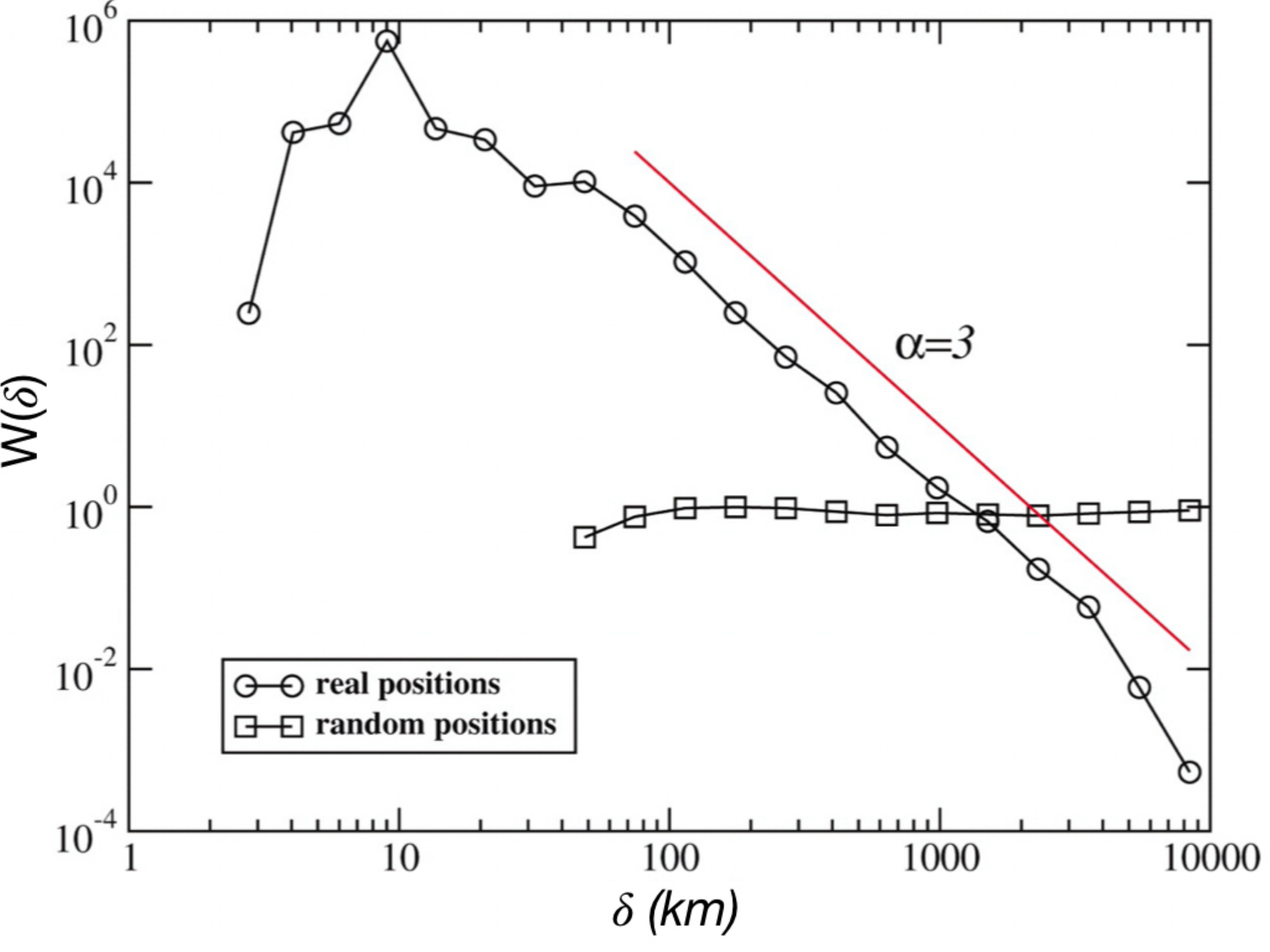}
\caption{The function $W(\delta)$ inferred by considering the canonical network model fixing the expected degree sequence and the expected distribution of distance $\delta$ between connected nodes, is plotted for the American Airport network \cite{colizza}. The figure show clearly that $W(\delta)$ displays a cutoff at small distances and at large distances decays as $W(\delta)\sim \delta^{-\alpha}$ with $\alpha\simeq 3$. Reprinted figure from Ref. \cite{PNAS}.}
\label{fig:American}
\end{figure}

\begin{figure}
      \centering 
\includegraphics[width=1.2\textwidth]{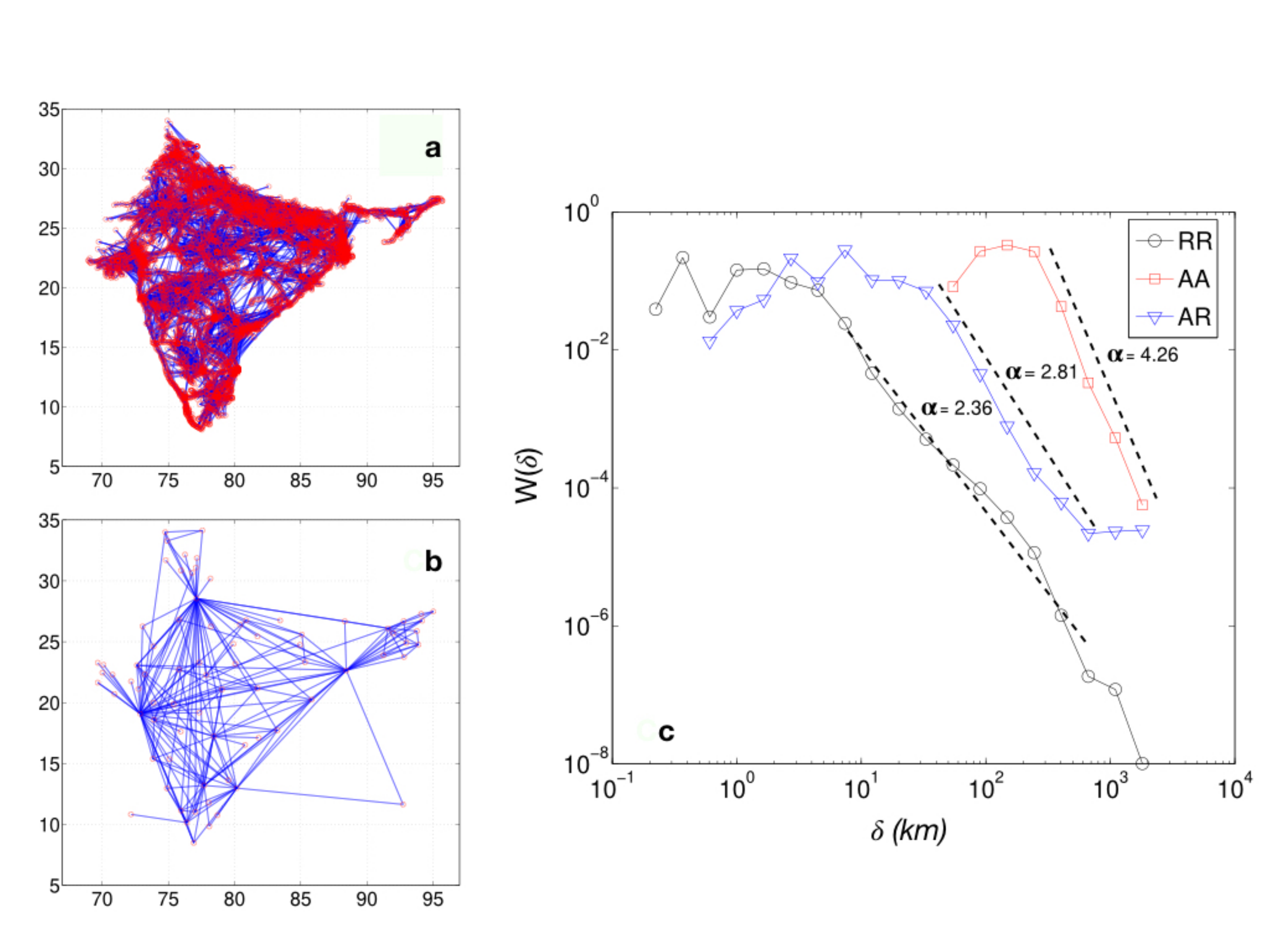}
\caption{The map of the Indian railway network (RR) and the Indian transportation network between railway stations and airports (AR) (panel a) and map of the  Indian airport network (AA) (panel b).Panel (c) displays the functions $W(\delta)$  for the Indian railway network (RR), Indian airport network (AA) and the bipartite network of interconnections between airports and train stations (AR) . At large distance the functions $W(\delta)$ for the three networks decay as a power-law of distance $W(\delta)\propto {\delta}^{-\alpha}$, with the value of the fitted exponents $\alpha$ indicated in the figure. Reprinted figure from \cite{Halu}. }
\label{fig:India}
\end{figure}

\subsection{Constraining the  distribution of distances between  linked nodes}
 \label{sec:spatial:dist}
 
In this paragraph we will discuss the   main bottom-up approach \cite{PNAS,bianconi2007entropy,bianconi2009}  to generate  suitable spatial network ensembles,  that can be used to randomize a given network and formulate  spatial null models.
To this end   let us  bin the distance $d({\bf x}_i,{\bf x}_j)$ between  any two nodes $i$ and $j$ in a vector of $Q$ distance intervals $\nu$.  We define the characteristic function $ \mathbbm{1}(d({\bf x}_i,{\bf x}_j),\delta_{\nu})$ indicating whether the distance between two nodes falls in a given distance interval $[\delta_{\nu},\delta_{\nu}+\Delta \delta)$ as
\bea
 \mathbbm{1}(d({\bf x}_i,{\bf x}_j),\delta_{\mu})=\left\{\begin{array}{ll}& 1  \ \mbox{if} \ d({\bf x}_i,{\bf x}_j)\in [\delta_{\mu},\delta_{\nu}+\Delta \delta),\\
&0 \ \mbox{otherwise}, \end{array}\right.
\eea
where we assume $1\leq \nu\leq Q$.
We consider the  canonical network ensemble in which we impose $\hat{P}=N+Q$ constraints: the set of canonical constraints enforcing the expected degree sequence and the set of canonical constraints enforcing the  expected number of links connecting nodes in a given distance interval, i.e.
\bea
\begin{array}{cclll}
{{k}_i}&=&\sum_{G\in \Omega_G} \left[P(G|{\bf x})F_i(G)\right],&\mbox{for}& 1\leq i\leq N,\nonumber \\
{{L}}(\delta_{\nu})&=&\sum_{G\in \Omega_G} \left[P(G|{\bf x})F_{\nu}(G))\right],&\mbox{for}& 1\leq\nu\leq Q,
\end{array}
\eea
with 
\bea
F_i(G)&=&\sum_{j=1}^N a_{ij},\nonumber \\
F_{\nu}(G)&=&\sum_{i<j}^N a_{ij} \mathbbm{1}(d({\bf x}_i,{\bf x}_j),\delta_{\nu}).
\eea
According to the general theory of canonical network ensembles, discussed in paragraph \ref{sec:canonical},
the probability of a spatial network 
in which we enforce  these constraints reads
\bea
P(G|{\bf x})=\frac{1}{Z}\exp\left[-\sum_{i=1}^N \lambda_i\sum_{j=1}^N a_{ij}-\sum_{\nu=1}^Qw_{d_{\nu}} \sum_{i<j}a_{ij}\mathbbm{1}(d({\bf x}_i,{\bf x}_j),\delta_{\nu})\right],
\eea
where $\lambda_i$ $w_{d_{\nu}}$ are the Lagrangian multipliers enforcing the constraints
\bea
{k}_i&=&\sum_{j=1}^N p_{ij}=\sum_{j=1}^N\sum_{\nu=1}^Q\frac{e^{-\lambda_i-\lambda_j-w_{\delta_{\nu}}}}{1+e^{-\lambda_i-\lambda_j-w_{\delta_{\nu}}}}\mathbbm{1}(d({\bf x}_i,{\bf x}_j),\delta_{\nu}),\\
{L}(\delta_{\nu})&=&\sum_{i<j}\frac{e^{-\lambda_i-\lambda_j-w_{\delta_{\nu}}}}{1+e^{-\lambda_i-\lambda_j-w_{\delta_{\nu}}}} \mathbbm{1}(d({\bf x}_i,{\bf x}_j),\delta_{\nu}).
\eea
The  marginal probability $p_{ij}$ a link  $(i,j)$ is given by
\bea
p_{ij}=\sum_{\nu=1}^Q\frac{e^{-\lambda_i-\lambda_j-w_{\delta_{\nu}}}}{1+e^{-\lambda_i-\lambda_j-w_{\delta_{\nu}}}} \mathbbm{1}(d({\bf x}_i,{\bf x}_j),\delta_{\nu}).
\label{pijs}
\eea  
By introducing the variables $\theta_i$ and $W(\delta_{\nu})$ defined as
\bea
\theta_{i}&=&e^{-\lambda_{i}},\\
W(\delta_{\nu})&=&e^{- w_{\delta_{\nu}}},\label{W2}
\eea
we can alternatively write the marginal probability $p_{ij}$ given by Eq. (\ref{pijs}) as
\bea
p_{ij}=\sum_{\nu=1}^Q\frac{\theta_{i}\theta_{j} W(\delta_{\nu})}{1+\theta_{i}\theta_{j}W(\delta_{\nu})}  \mathbbm{1}(d({\bf x}_i,{\bf x}_j),\delta_{\nu}).
\label{W}
\eea
From this expression it is clear that the dependence of the marginal with the distance between the two connected nodes is fully captures by the function of $W(\delta)$.The function $W(\delta)$ can be directly inferred from data by numerically solving finding the Lagrangian multipliers and using  Eq.(\ref{W2}). As it can be seen from the expression of the marginal $p_{ij}$ given by Eq. (\ref{W}). $W(\delta)$ is fixed modulo a constant by Eq. (\ref{W}) and describes how the the marginal probability of a link depend on the distance between the two connected nodes.
In Figure \ref{fig:American} we show the function $W(\delta)$ find in Ref.\cite{PNAS} for the American Airport network \cite{colizza}.
We show that $W(\delta)$ feature a cutoff at small distances indicating that flights must connect distant enough locations.  However for large distances $W(\delta)$  it display a clear power-law dependence on the distance $d$, $W(\delta)\sim d^{-\alpha}$ with $\alpha\simeq 3$.
Clearly, if instead of taking the real positions of airport in space we consider the same set of node positions but we assign these position to randomly reshuffled node labels, the function $W(\delta)$ is well approximated by a constant (see Figure \ref{fig:American}).  Interestingly this power-law behavior of the function $W(\delta)$ is also displayed by the different layers of the multilayer spatial transportation network \cite{bianconi2018multilayer} in India \cite{Halu}. In  Figure \ref{fig:India} we show  the function $W(\delta)$  for the network between railway stations (RR) the bus transportation network  between railway stations  and airports (RA) and the Indian airport network (AA). All networks display a   power-law behaviour of the function $W(\delta)\simeq \delta^{-\alpha}$ although with different exponents $\alpha$. 

\begin{figure}
\centering 
\includegraphics[width=0.95\textwidth]{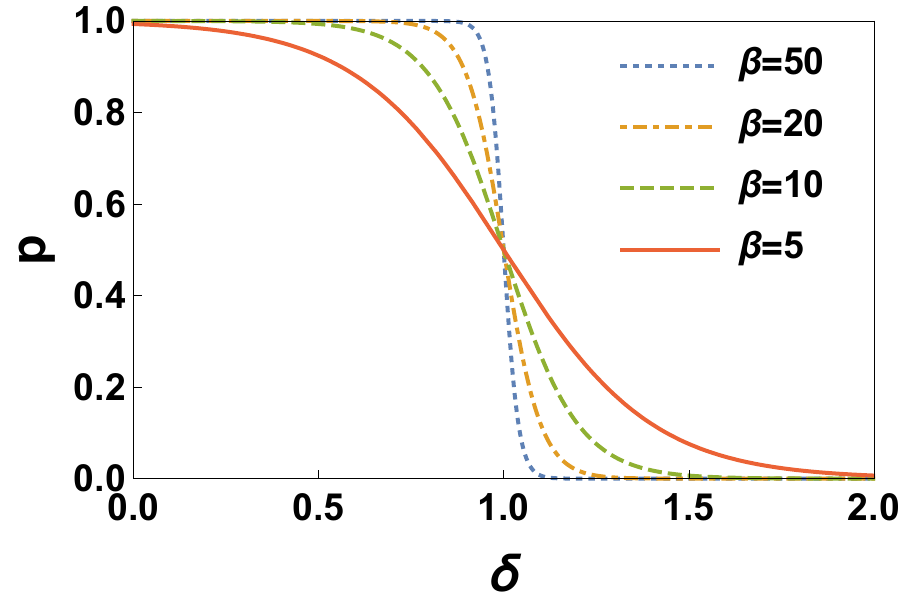}
\caption{
Functional form of the marginal probability $p$ of a link at distance $\delta$ in a soft random geometric ensembles as defined given by Eq. (\ref{eq:Fermi}). The different curves refers to different values of $\beta$ indicated in the legend, with $r_0=1$. As $\beta\to \infty$  the distribution approaches a step function indicating that nodes are connected with probability one if and only if they are at distance $\delta<r_0$. Therefore in the limit $\beta\to\infty$ the network in this ensemble reduce to  random geometric networks.
}
\label{fig:fermi}
\end{figure}
    \subsection{Global spatial constraints}
    \subsubsection{Spatial network ensemble with global spatial constraint}
    In the previous paragraph we have  taken a bottom-up approach and we have seen how to construct a spatial networks in which the distribution of distances between linked nodes is preserved in expectation. 
    Here we take a different, top-down approach in which we consider a  given function $f({\bf x}_i,{\bf x}_j)$
indicating the “cost” of a link connecting  node with latent variable  ${\bf x}_i$
and a node with latent variable ${\bf x}_j$.
This function might also be attributed a  different meaning and depending on the case under study we can  give to  it the meaning of “cost” or instead the meaning of  “payoff/benefit”.

We therefore consider the canonical spatial ensemble in which we fix just two constraints, i.e. $\hat{P}=2$. The first constraint is the expected total number of links and the second constraint is the expected cost of all the links of the network.
These canonical constraints are of the type defined in Eq. (\ref{sp:const}) and read 
\bea
{{L}}&=&\sum_{G\in \Omega_G} P(G)F_1(G),\nonumber \\
{C}&=&\sum_{G\in \Omega_G} P(G)F_2(G),
\eea 
with $F_1(G)$ and $F_2(G)$ given by 
\bea
F_1(G)&=&\sum_{i<j}^N a_{ij},\label{cs1}\\
F_2(G)&=&\sum_{i<j}^N a_{ij}f({\bf x}_i,{\bf x}_j).\label{cs2}
\eea
Applying the general theory of canonical network ensembles described in paragraph \ref{sec:canonical} it is immediate to derive the Gibbs measure of this ensemble given by
\bea
P(G|{\bf x})=\frac{1}{Z}\exp\left[-\beta \sum_{i<j}a_{ij} f({\bf x}_i,{\bf x}_j)-\hat{\mu}\sum_{i<j}a_{ij}\right],
\eea
where $\beta $ and $\hat{\mu}$ the Lagrangian multipliers fixed by the  constraints specified in Eq.(\ref{cs2}) and Eq.(\ref{cs1}).
Moreover the marginal probability $p_{ij}$ that two nodes 
$i$ with latent variables ${\bf x}_i$
and 
$j$ with latent variables ${\bf x}_j$
are linked 
is only a function of the latent variables 
and is given by 
\bea
p_{ij}=p({\bf x}_i,{\bf x}_j)=\frac{e^{-\beta f({\bf x}_i,{\bf x}_j)-\hat{\mu}}}{1+e^{-\beta f({\bf x}_i,{\bf x}_j)-\hat{\mu}}}.
\label{sp:glo:pij}
\eea
As we will see in the next paragraphs, this ensemble includes maximum entropy ensembles of very different natures spanning from ensemble of networks with latent variables introduced in Ref. \cite{caldarelli2002scale} and constituting an interpretation of the canonical network model with expected degree sequence to networks which are totally driven by the underlying geometry such as  random geometric networks \cite{walters2011random}. As we will see this class of models includes also the random hyperbolic network model investigated in  Ref. \cite{krioukov2010hyperbolic}. 
\subsubsection{Scalar latent variables}
Let us know consider the ensemble just discussed and consider the simple case in which the latent variable ${\bf x}_i$ associated to a given node $i$ is a scalar and let us call the scalar $x_i$. This ensemble has been proposed as a model for complex networks in Ref.\cite{caldarelli2002scale}. This model can be used by taking the scalar variables $x_i$ from some meta-data associate to the nodes, such as the GDP of a country in the World Trade networks \cite{garlaschelli2005structure}.
Given that  the latent variable is a scalar the marginal probability $p_{ij}$ given by Eq.(\ref{sp:glo:pij}) reads
\bea
p_{ij}=p({ x}_i,{ x}_j)=\frac{e^{-\beta f({ x}_i,{ x}_j)-\hat{\mu}}}{1+e^{-\beta f({ x}_i,{ x}_j)-\hat{\mu}}}.
\eea
Therefore necessarily  all the nodes with latent variable $x_i=x$  have the same expected degree given by 
\bea
{k}_i=N\int dx \rho(x) p(x_i,x).
\eea
where $\rho(x)$ is the density distribution function of the latent variables  $x$. It follows that that  the latent variable is a proxy for the degree.
If we assume that the cost function $f(x_i,x_j)$ has the form
\bea
\beta f(x_i,x_j)=\lambda_{x_i}+\lambda_{x_j}
\eea
where $\lambda_{x}$ is a function of $x$, (ideally a one-to-one function of $x$) we have  that the marginal probability $p({x_i,x_j})$ reads
\bea
p({x_i,x_j})=\frac{e^{- (\lambda_{x_i}+\lambda_{x_j})-\hat{\mu}}}{1+e^{-(\lambda_{x_i}+\lambda_{x_j})-\hat{\mu}}}=\frac{\theta_{x_i}\theta_{x_j}/z}{1+\theta_{x_i}\theta_{x_j}/z}
\eea
where $\theta_{x_i}=e^{-\lambda_{x_i}}$ fixes the expected degree of node $i$ and  $z=e^{\beta\hat{\mu}}$ fixes the expected total number of links.
It follows that this ensemble upon identification of $\lambda_{x_i}$ with the value of the Lagrangian multipliers $\lambda_i$ reduces to the canonical ensemble with given expected degree sequence with marginal given by Eq. (\ref{pij}).
Therefore this ensemble relates  the scalar latent variable coming from meta-data associated to the nodes of the network to the expected degree of the nodes in the canonical ensemble with marginal given by Eq. (\ref{pij}).
\subsubsection{Spatial networks with given cost of the links at given distance}
\label{sec:spatial:cost}

Let us know introduce a cost function $f({\bf x}_i,{\bf x}_i)$  making use of the distance between the positions (or latent variables) of the nodes.
In order to have an ensemble that combines effect of the metric with terms that can control the expected degree sequence we consider the cost function 
\bea
\beta f({\bf x}_i,{\bf x}_j)=\lambda_{{\bf x}_i}+\lambda_{{\bf x}_j}+\beta {w}(d({\bf x}_i,{\bf x}_j)),
\eea
where $\delta_{ij}=d({\bf x}_i,{\bf x}_j)$ indicates the distance between node $i$ and node $j$ in the embedding space and $w(\delta)$ indicates the cost of links connecting nodes at distance $\delta$. Moreover the functions $\lambda_{{\bf x}}$ are only functions of the latent variable of one node and as we will see, as in the previous case, they fix the expected degree of the nodes.

In this case the marginal probability reads
\bea
p({{\bf x}_i,{\bf x}_j})=\frac{e^{- (\lambda_{{\bf x}_i}+\lambda_{{\bf x}_j})-\hat{\mu}-\beta {w}(\delta_{ij})}}{1+e^{-(\lambda_{{\bf x}_i}+\lambda_{{\bf x}_j})-\hat{\mu}-\beta {w}(\delta_{ij})}}=\frac{\theta_{{\bf x}_i}\theta_{{\bf x}_j} W(\delta_{ij})/z}{1+\theta_{{\bf x}_i}\theta_{{\bf x}_j}W(\delta_{ij})/z},
\eea
where we have defined 
\bea
\theta_{{\bf x}_i}&=&e^{-\lambda_{x_i}},\nonumber\\
z&=&e^{\hat{\mu}},\nonumber \\
W(\delta_{ij})&=&e^{-\beta w(\delta_{ij})}.
\label{sp:glo:W}
\eea

The function $w(\delta)$ indicating the cost  of a link connecting nodes at distance $\delta$  can depend on the distance in different ways. Two choices are particularly relevant to model data, by making minimal assumptions.
\begin{itemize}
\item {\em Cost  $w(\delta)$ linear in $\delta$.}\\
{If} the cost of a link $w(\delta)$ increases linearly with the distance $\delta$ of the two connected nodes, i.e.
\bea
{w(\delta)}= \delta,\eea
then the function $W(\delta)$ determining the dependence of the link probability on the distance $\delta$ of its two endnodes  decays exponentially with $\delta$, i.e.
\bea
 W(\delta)=e^{-\beta \delta}.\eea
\item
{\em Cost  $w(\delta)$ linear in $\ln \delta$.}\\
{If} the cost of a link $w(\delta)$ increases linearly with the order of magnitude of the distance $\delta$ of the two connected nodes, i.e.
\bea
{w(\delta)}=\ln \delta,
\eea
then the function $W(\delta)$ determining the dependence of the link probability on the distance $\delta$ of its two endnodes  decays as a power-law of  $\delta$, i.e.
\bea W(\delta)=\delta^{-\beta}. \eea

\end{itemize}

We are now in the position to give a further interpretation of the observed power-law decay of $W(\delta)$ 
for the American airport network and the Indian railway/airport network (see Figure $\ref{fig:American}$, and Figure \ref{fig:India}). 
Indeed we  can  interpret this result as the outcome 
of an effective “cost” of the connections 
growing proportionally 
to the order of magnitude of distance between the nodes.
\subsubsection{Soft random geometric networks }

The ensemble of spatial networks attributing to each link a given cost can capture also soft random geometric networks.
This limit is achieved if we take the cost function
\bea
f({\bf x}_i,{\bf x}_j)=d({\bf x}_i,{\bf x}_j)-r_0,
\eea
i.e. we give a positive cost to links connecting nodes at distance greater than $r_0$ and a negative cost to links connecting nodes at distance  smaller than $r_0$.
With this cost function  the marginal $p_{ij}$ given in general by  Eq.(\ref{sp:glo:pij}) depends only on $\delta_{ij}=d({\bf x}_i,{\bf x}_j)$ and takes the form,
\bea
p_{ij}=p(\delta_{ij})=\frac{e^{-\beta (\delta_{ij}-r_0)-\hat{\mu}}}{1+e^{-\beta (\delta_{ij}-r_0)-\hat{\mu}}}=\frac{1}{e^{\beta(\delta_{ij}-r_0)+\hat{\mu}}+1}
\label{eq:Fermi}
\eea
which  is shown in Figure \ref{fig:fermi} for different values of $\beta$.

For $\hat{\mu}=0$ and $\beta\to\infty$ the marginal reduces to step function and 
two nodes
 at distance $\delta$
are connected
 if and only if  $\delta\leq r_0$. Therefore in this limit this ensemble reduces to a random geometric network \cite{walters2011random} where all pairs of nodes at distance is $\delta_{ij}<r_0$ are connected. If $\beta$ is finite the step function is smoothed out allowing links at distance $\delta>r_0$ with a small probability, therefore for finite value of $\beta$ the networks in this  ensemble are called soft random geometric networks. Finally for $\beta=0$ the marginal is independent of the distance between the nodes.

\subsubsection{Hyperbolic random geometric networks}

The random geometric network is a framework that can be applied to build spatial network starting from any distribution of nodes in space and any metric distance defined on the embedding space. A particular attention has been recently addressed to random geometry graphs in hyperbolic space  \cite{krioukov2010hyperbolic,boguna_navigability,Serrano_trade,boguna2010sustaining,Embedding,garcia2019mercator}  to which we devote  this  paragraph.

\begin{figure}
\begin{center}
\includegraphics[width=0.7\textwidth]{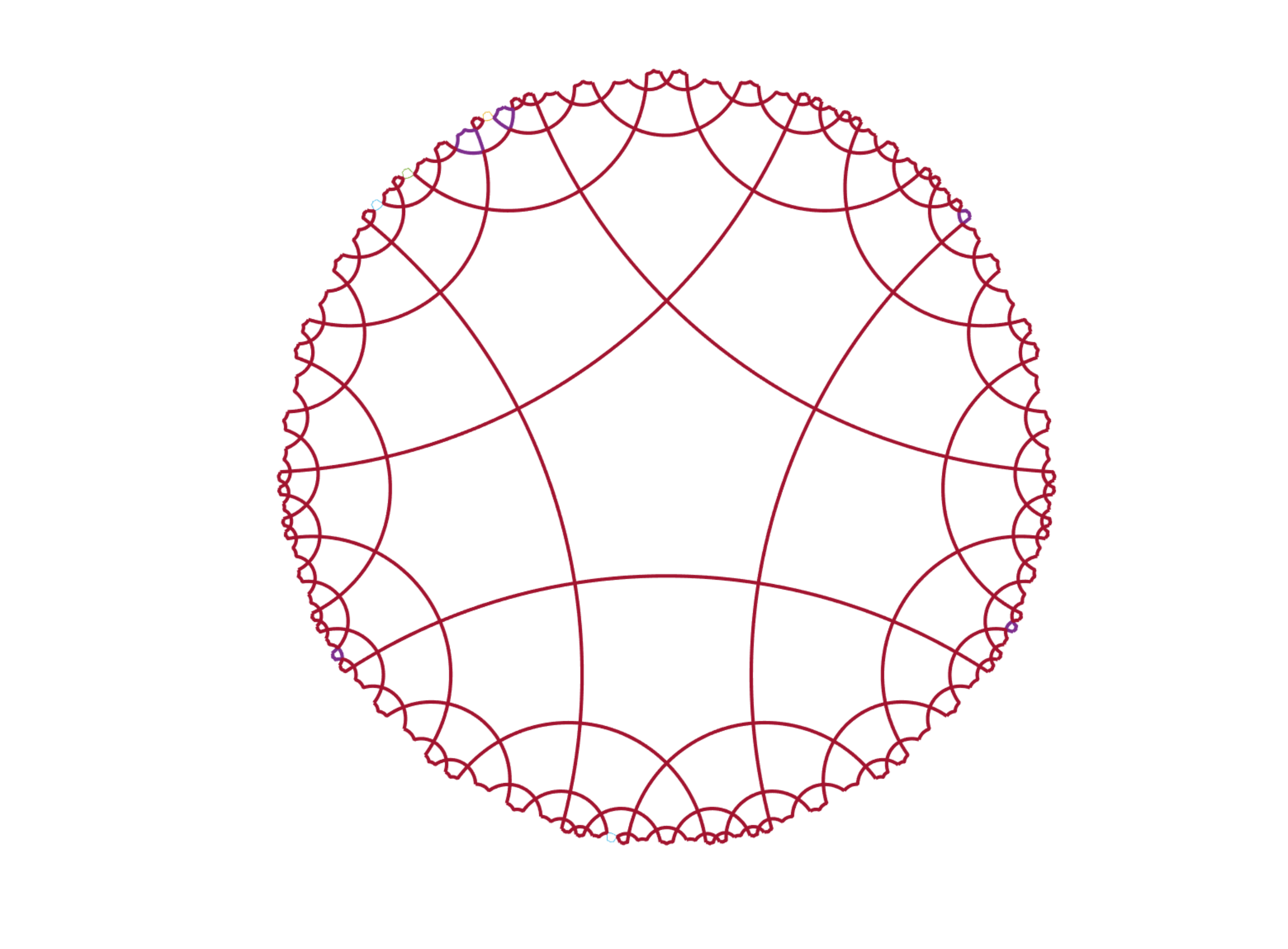}
\caption{A lattice of the hyperbolic space $\mathbbm{H}^2$ where geodesic among neighbour nodes are indicated.}
\label{fig:hyperbolic}
\end{center}
\end{figure}

Let us consider  unitary ball $\mathcal{B}=\{{\bf x}\in \mathbbm{R}^2||{\bf x}|<1\}$. To each pair of nodes $i$ and $j$ of polar coordinates $(r_i,\phi_i)$, $(r_j,\phi_j)$      we associate a distance            
$\delta_{ij}=d({\bf x}_i,{\bf x}_j)$ satisfying  
\bea
\cosh \zeta \delta_{ij}= \cosh \zeta r_i \cosh\zeta r_j -\sinh \zeta r_i\sinh \zeta r_j \cos \Delta \phi_{ij},
\label{hyp_d1}
\eea
where $\Delta \phi_{ij}=\pi -|\pi-|\phi_i-\phi_j||$ is the angular distance between the nodes.
The distance defined in Eq. (\ref{hyp_d1}) reduces the set of all position vectors in the ball $\mathcal{B}$ into  the model 
$\mathbbm{H}^2$ of hyperbolic space of constant curvature $R=-\zeta^2$ (see Figure $\ref{fig:hyperbolic}$).

We can use this distance to define a hyperbolic random geometry network which is a maximum entropy ensemble where each link contributes to the global cost function  by a term
 \bea
 f({\bf x}_i,{\bf x}_j)=\zeta(\delta_{ij}-r_0)\simeq \zeta r_i + \zeta r_j + \ln{\Delta \phi_{ij}}-r_0
 \eea
where, following Ref. \cite{krioukov2010hyperbolic} we have approximated by expression for the hyperbolic distance 
given by Eq. (\ref{hyp_d1}) by
\bea
 \delta_{ij}\simeq r_i + r_j +\frac{2}{\zeta} \ln \sin \frac{\Delta \phi_{ij}}{2}\simeq r_i + r_j +\frac{1}{\zeta} \ln {\Delta \phi_{ij}}.
 \label{dij:hyp}
\eea

Given the general solution of random geometric networks, by   absorbing the Lagrangian mutlipler $\hat{\mu}$  into the parameter $r_0$, i.e.  putting $\hat{\mu}=0$,
the marginal probability reads
\bea
p_{ij}=\frac{e^{-\beta \zeta (\delta_{ij}-{r}_0)}}{1+e^{-\beta \zeta (\delta_{ij}-{r}_0)}}=\frac{1}{e^{\beta\zeta (\delta_{ij}-{r}_0)}+1}
\eea
where $r_0$ has the  role to  tune the number of the links in the ensemble
 and where $\delta_{ij}$ is given in Eq.(\ref{dij:hyp}).
This ensemble is equivalent to an ensemble discussed in paragraph \ref{sec:spatial:cost} with marginal given by 
\bea
p_{ij}=\frac{\theta_i\theta_jW(\Delta\phi_{ij})}{1+\theta_i\theta_j W(\Delta \phi_{ij})}
\label{marginal:hyp}
\eea 
in which we  put
\bea
\theta_{i}&=&e^{-\beta \zeta (r_i-r_0/2)},\nonumber\\
W(\Delta \phi_{ij})&=&e^{-\beta\ln \Delta \phi_{ij}}=(\Delta \phi_{ij})^{-\beta}=(\Delta \phi_{ij})^{-\beta}.
\label{thetah}
\eea
Therefore the random geometric hyperbolic network model reduces 
to a maximum entropy model in which we fix the expected  degrees 
and we have a cost of the link proportional to the 
order of magnitude of the angular distance between the nodes.
It follows that  the hyperbolic random geometric networks can be understood as a way to have  a global geometrical description that merges variables not associated to a metric distance (associated to the radial coordinate in hyperbolic space) to variables  associated to a metric distance (the angular coordinate of nodes in the hyperbolic disk).
The only difference between the random geometric hyperbolic network and the maximum entropy ensemble with cost of the links is that the degrees of the nodes are fixed by the embedding of the nodes in space (specifically by their radial coordinate). Indeed the expected degree ${k}_i$ of node $i$ is given by 
\bea
{k}_i=\sum_{j=1}^Np_{ij}=\sum_{j=1}^N\frac{\theta_i\theta_j(\Delta\phi_{ij})^{-\beta}}{1+\theta_i\theta_j (\Delta \phi_{ij})^{-\beta}}.
\label{mh}
\eea
Therefore the degree distribution is fixed by the distribution of nodes in space.
Starting from the marginal given by Eq.(\ref{marginal:hyp}) where $\theta_i$ are expressed in term of the radial coordinate $r_i$ as prescribed by Eq.(\ref{thetah}) we observe that  in the sparse regime we have that the degree $k$ of a node is related to its radial coordinate by $k\propto \theta$ and specifically 
\bea
k=A e^{-\beta \zeta r}.
\label{krchange}
\eea
In other words,  the radial distribution of the nodes determines the degree distribution. In particular the radial distribution $\rho(r)$ of the nodes  that enforces a power-law degree distribution $P(k)\simeq Ck^{-\gamma}$   with exponent      $\gamma\geq 2$, 
 is exponential  and given by 
\bea
\rho(r) \propto e^{\beta \zeta(\gamma-1) r}.
\eea
To show this it is sufficient to use Eq.(\ref{krchange}) to do a change of variables. In particular using 
\bea
\rho(r) = P(k)\left|\frac{dk}{dr}\right|,
\eea
we get 
\bea
\rho(r) = P(k)\left|\frac{dk}{dr}\right|_{k=Ae^{-\beta \zeta r}}=\left. AC \beta\zeta k^{-\gamma} e^{-\beta \zeta r}\right|_{k=Ae^{-\beta \zeta r}}=AC\beta \zeta e^{\beta \zeta(\gamma-1) r}.
\eea
It follows that a hyperbolic random geometric network, in the sparse regime   generates scale free networks if the radial distribution of nodes is exponential.

This model has been used to infer the hidden geometry of several real datasets in Refs. \cite{boguna2010sustaining,Serrano_trade} including the Internet at the Autonomous System Level and the Trade World Networks. Moreover the embedding of these networks in hyperbolic space has been use to propose a greedy algorithm to efficiently navigate the networks \cite{boguna_navigability}. Recently this ensembles of networks has also be related to maximum entropy ensemble with non-trivial block structure (the so called block models)  \cite{faqeeh2018characterizing}.
This model has a non equilibrium (growing) network counterpart proposed in \cite{papadopoulos2012popularity} and has boosted the research on finding efficient embedding algorithms of networks in hyperbolic space \cite{garcia2019mercator,Embedding}.
 The network generated by the discussed series of works are never discrete manifolds, and assume that the establishment of links depends on the underlying hyperbolic distance. An alternative framework called ``Network Geometry with Flavor" \cite{bianconi2017emergent,bianconi2016network,mulder2018network} instead is able to generate hyperbolic manifolds of any dimension $D$ by a mechanism of emergent network geometry. In this case, discrete manifolds that tile the hyperbolic spaces are generated by combinatoric rules that make no use of the underlying geometry of the network \cite{bianconi2017emergent,mulder2018network}.

\section{Classical information theory of networks}

\subsection{Classical network ensembles and optimal input distribution}

In this section we summarize the most salient aspects of the recently proposed classical information theory of networks\cite{radicchi2015percolation}. 

The classical information theory of networks is based on a classical network ensemble and allows us to address a fundamental question in network theory:{\em which is the optimal distribution of  the latent variables fixing the network constraints?}

Until now we have discussed maximum entropy network ensembles satisfying a given set of constraints but we have not discussed where these constraints come from. However in network theory there is evidence that networks follow universal statistical properties, shared by a vast number of networks having very different functions. For instance  the scale-free degree distribution \cite{barabasi1999emergence} is a property shared by different systems ranging from the Internet to collaboration networks, and to biological networks as well. Additionally, in spatial  spatial networks the nodes' positions are in most of the cases not distributed  uniformly in space, as it can be see from the worldwide distribution of airports or railway stations.  The evidence that real data is not uniformly distributed is also called the ``blessing of non-uniformity" of data, a feature that is known to allow for  a  better efficiency of inference algorithms \cite{domingos2012few}.

 Here we will show how information theory can allow us to define a principle for determining the optimal distribution of latent variables determining the network constraints. We call this distribution the {\em optimal latent variables distribution}.

The optimal latent variables distribution is defined in the framework of the classical information theory of networks \cite{radicchi2019classical} making use of classical network ensembles. In the classical network ensemble we abandon the description of the network ensemble as a probability measure over adjacency matrices and we consider instead the network ensemble as a probability measure over the edge list, i.e. we consider the  links of a networks as independent random variables given by the labels if their two endnodes. Interestingly in this way the classical network ensemble reconciles maximum entropy models for networks  with the well known maximum entropy models for urban  mobility. \cite{WILSON1967253,wilson1969use,wilson2011entropy}.

If follows that the classical network ensembles can be interpreted as source of information whose  messages are the links of the network.
 The optimal latent variables distribution  is obtained by considering a lossy compression channel that instead of transmitting the original messages coding for the network (the list of the label of the two endnodes of each link) transmit exclusively the latent variables associated to the connected nodes. Therefore this lossy compression channel removes the information about the identity (labels of the nodes) and treats all the pair of nodes with the same latent variables on the same footing.
 While the original channel has entropy $S$ the lossy compression channel has entropy $H$. The optimal input distribution, i.e. the optimal distribution of latent variables associated to links is the distribution that maximizes the entropy $H$ given a fixed value of the entropy $S$, i.e. maximizes the entropy of the lossy compression channel acting on classical network ensembles at fixed information content of the classical network ensemble.
 
The optimal latent variables distribution can take the form of optimal degree distribution if we consider the classical network ensemble with given expected degree sequence; alternatively for spatial networks the optimal latent variables distribution can define an optimal distribution of distances between pair of nodes. Interesting we found that the optimal latent variable distribution are highly heterogeneous.In the first case in which the only latent variable si the expected degree of the nodes the optimal latent variables distribution  reduce to the scale-free degree distribution. in the second case in which the latent variable is the distance between the nodes in their latent space, the optimal latent variable distributions reduces to  a highly inhomogeneous (and eventually fractal) distribution of nodes in spatial networks.  

These results can be used to provide an information theory explanation of the observation of universal properties of complex network. Moreover the optimal latent variables distribution are  important candidates for priors over latent variables to be used in inference problems.

\subsection{Classical network ensembles}

\subsubsection{The entropy of classical network ensembles}
The classical network ensemble  \cite{radicchi2015percolation} is based on the edge-list representation of a network where each link it is considered as an independent variable that has value given by the labels of its two endnodes. Therefore this approach reconciles maximum entropy network models with maximum entropy models for  urban mobility \cite{WILSON1967253,wilson1969use,wilson2011entropy}.
As we will see in the next chapter this framework is very suitable to define the concept of {\em optimal latent variables distribution}.

In the classical network ensemble we consider the set $\Omega_G$  of all  networks $G=(V,E)$ of given total number of nodes $N$ and given total number of links $L$, i.e. with $|V|=N$  and $|E|=L$.

Each network is specified by the list of  $L$ labelled links with each link $n$   denoted as  
\bea
\vec{\ell}^{[n]}=\left(\ell_1^{[n]},\ell_2^{[n]}\right),
\eea 
with $1\leq n\leq L$. Here  $\ell_1^{[n]}$ and $\ell_2^{[n]}$  indicate the label of the first endnode and of  the second endnode of the $n$-th link respectively.  T

The classical network ensemble is the  ensemble in which each network $G$  is associated a probability $P(G)$ given by 
\bea
{P}\left(G\right)=\prod_{n=1}^L \tilde{P}\left(\vec{\ell}^{[n]}\right)
\eea
where $\tilde{P}\left(\vec{\ell}^{[n]}\right)$ is the probability that a random link has endnodes specified by the vector $\vec{\ell}^{[n]}$.
In other words the network is the results of $L$ independent drawing of the links from the distribution $\tilde{P}(\vec{\ell}).$
The classical network ensemble has therefore entropy given by 
\bea
S(G)=-L\sum_{\vec{\ell}}\tilde{P}(\vec{\ell})\ln \tilde{P}(\vec{\ell}).
\label{Sc}
\eea

Let us assume that the probability that the two endnodes of a randon link are node $i$ and node $j$ only depends on their latent variables, i.e.
\bea
\tilde{P}(\vec{\ell}=(i,j))=\pi({\bf x}_i,{\bf x}_j)
\eea
This is a very general expression, and it includes various relevant cases.
For instance if  the latent variable indicates  the expected degrees we will have 
\bea
\pi({\bf x}_i,{\bf x}_j)=\pi({ k}_i,{k}_j).
\eea
In the case in which  the latent variables indicate the position of the nodes in space we can describe the case in which the $\pi({\bf x}_i,{\bf x}_j)$ depends on the distance of the nodes as
\bea
\pi({\bf x}_i,{\bf x}_j)=W(d({\bf x}_i,{\bf x}_j)).
\eea
In that case the classical entropy $S(G)$ takes the simple form 
\bea
S(G)=-L\sum_{i,j}\pi({\bf x}_i,{\bf x}_j)\ln \pi({{\bf x}_i,{\bf x}_j}).
\eea
Let us  indicate with  $ P_{\mathcal V} ({\bf X})$ the fraction  of pair of nodes with latent variable ${\bf X}={\bf X}_{ij}=({\bf x}_i,{\bf x}_j)$, with 
\bea
P_{\mathcal V} ({\bf X})=\frac{1}{N^2}\sum_{i,j}\tilde{\delta}({\bf
   X}_{ij},{\bf{X}}),
   \eea 
   where $\tilde{\delta}(x',y')$ denotes the delta function. 
   We observe that we   can express the classical entropy $S$ in terms of the distribution $P_{\mathcal V} ({\bf X})$ as    
\bea
 S=-L\sum_{{\bf X}}N^2P_{\mathcal V}({\bf X})\pi({\bf{X}})\ln \pi({\bf X}).
 \label{Scond}
 \eea   
\subsubsection{Maximum entropy classical ensembles}
   The maximum entropy classical network ensembles are the ensembles that maximize the classical entropy $S(G)$ defined in Eq.(\ref{Sc}) under a specified set of constraints.
Let us indicate the general constraints as 
   \bea
\sum_{\vec{\ell}}{P}(\vec{\ell})F_{\mu}(\vec{\ell})=C_{\mu},
\label{lcon}
\eea
with $1\leq \mu\leq \hat{P}$.
Here $F_{\mu}(\vec{\ell})$ is an observable associated to a link which might indicate for instance if the link is connected to a specified node.
By maximizing the entropy $S(G)$ defined in Eq.(\ref{Sc}) under these constraints we get that the maximum entropy distribution $\tilde{P}(\vec{\ell})$ is given by the Gibbs measure
\bea
\tilde{P}(\vec{\ell})=\frac{1}{Z}e^{-\sum_{\mu=1}^P\lambda_{\mu}F_{\mu}(\vec{\ell})},
\label{solc}
\eea
where $Z$ is the normalization sum and $\lambda_{\mu}$ are the Lagrangian multipliers enforcing the constraints specified by Eq. (\ref{lcon}).

To be concrete we can consider the maximum entropy classical network ensemble in which we fix $N$ constraints of the type defined in Eq.(\ref{lcon}) with observables $F_i$ and constraints $C_i$ given by 
\bea
F_i(\vec{\ell})&=&\hat{\delta}_{\ell_1,i}+\hat{\delta}_{\ell_2,i},\nonumber \\
C_i&=&\frac{k_i}{L},
\label{Fi}
\eea
with $1\leq i\leq N$. Each  constraint $i$ imposes that the expected number of links incident to the generic node $i$ is given by  $k_i$.
Using the general solution Eq.(\ref{solc}) and imposing the mentioned constraints it is straightforward to see that the marginal $\pi_{ij}$ can be expressed simply as
\bea
\pi_{ij}=\pi(k_i,k_j)=\frac{k_ik_j}{(\langle k\rangle N)^2},
\label{cla:pikk'}
\eea
where $L=\avg{k}N/2$.
Therefore the expected number of links between node $i$ and node $j$ is given by
\bea
\avg{a_{ij}}=2L\pi_{ij}=\frac{k_i k_j}{\langle k\rangle N}.
\eea
This result implies that in the classical network model we do not get a marginal that follows the Fermi-Dirac distribution, as obtained in paragraph \ref{sec:can:ex} for the canonical network ensemble in which we fix the expected degree sequence; rather  we get directly the expression of the marginal of an  uncorrelated network ensemble. Note however that since the classical network ensemble allows for multiple links (i.e. pair of nodes connected by more than one link) for broad degree distributions that do not display the structural cutoff we will have $\avg{a_{ij}}>1$ at least for some links of the network.
The classical entropy of this ensemble directly as a function of the degree distribution $P(k)$  and is given by 
\bea
S=\langle k\rangle N\left[\ln (\langle k\rangle N)-\sum_{k}\frac{kP(k)}{\langle k \rangle}\ln k\right].
\eea

The second example we consider is the maximum entropy classical network  ensembles in which we impose a single global constraint but this constraint depends on the position of node in some embedding metric space where a distance between the nodes is defined.
We therefore consider a single $\hat{P}=1$ constraint or type Eq.(\ref{lcon}) associated to the observable
\bea
F(\vec{\ell})=\sum_{i<j}w(\delta_{ij})
\label{lf}
\eea
where $w(\delta_{ij})$ is an arbitrary ``cost" function of the distance $\delta_{ij}$ between two nodes.
In this case the marginal can be expressed as  
\bea
\pi_{ij}=\pi_{ij}(\delta_{ij})=\frac{g(\delta_{ij})}{N^2},
\label{cla:pidelta}
\eea
where 
\bea
g(\delta)=\frac{e^{-\beta w(\delta)}}{z},
\eea
and $z$ is the normalization constant ensuring $\sum_{ij}\pi_{ij}=1$.
By defining $\omega(\delta)$ as the probability desity distribution that two random nodes are at distance $\delta$ the entropy of this spatial network ensemble can be expressed as 
\bea
S=\langle k\rangle N\ln N-\frac{1}{2}\langle k\rangle N\int d\delta \ \omega(\delta)g(\delta) \ln g(\delta).
 \eea
 Finally  we consider a third example of classical network ensemble satisfying  $\hat{P}=N+1$ constraints given by Eq. (\ref{lcon}), with $N$ constraints (given by Eq.(\ref{Fi})) fixing the expected number of links incident to each node of the network and one global constraint defined by the observable Eq.(\ref{lf}) fixing the cost associated to sustaining connection of nodes at distance $\delta$.
 In this case the marginal distribution reads 
 \bea
 \pi_{ij}=\pi(\theta_i,\theta_j,\delta_{ij})=\frac{\theta_i\theta_j g(\delta_{ij})}{N^2},
 \label{sp:m:2}
 \eea
 where we have defined 
 \bea
 \theta_i&=&e^{-\lambda_i},\nonumber \\
 g(\delta_{ij})&=&\frac{e^{-\beta w(\delta_{ij})}}{z},
 \eea 
 where $\lambda_i$ and $\beta$ are the Lagrangian multiplies that fix the expected degree of node $i$, and  the expected value of the cost function given by Eq.(\ref{lf}) respectively and where  $z$ is the normalization constant, enforcing $\sum_{i,j}\pi_{ij}=1$.
 The entropy of this ensemble reads
 \bea
S&=&-L\sum_{ij}\pi_{ij}\ln \pi_{ij}=\avg{k}N\ln N  \nonumber \\
&&\hspace*{-5mm}-\frac{1}{2}\avg{k}N\int d\theta\int d\theta'\int d{\delta} \, {\omega}(\theta,\theta',\delta)\,\theta\theta'g(\delta)\ln [\theta\theta'g(\delta)],
\eea
where ${\omega}(\theta,\theta',\delta)$ is the probability density of pairs of nodes with latent variables $\theta$ and $\theta'$ at distance $\delta$.
\subsection{Optimal latent distribution}
\subsubsection{Optimization principle}
 A classical network ensemble describes a network as a set of $L$ links where each link $\vec{\ell}$ is an independent random variable drawn from the distribution $\tilde{P}(\vec{\ell})$. Therefore each network can be considered as a
 source of $L$ messages where each message is  a link $\vec{\ell}$ specified by the ordered pair of its two endnodes.
 
\begin{figure}
\begin{center}
\includegraphics[width=0.6\textwidth]{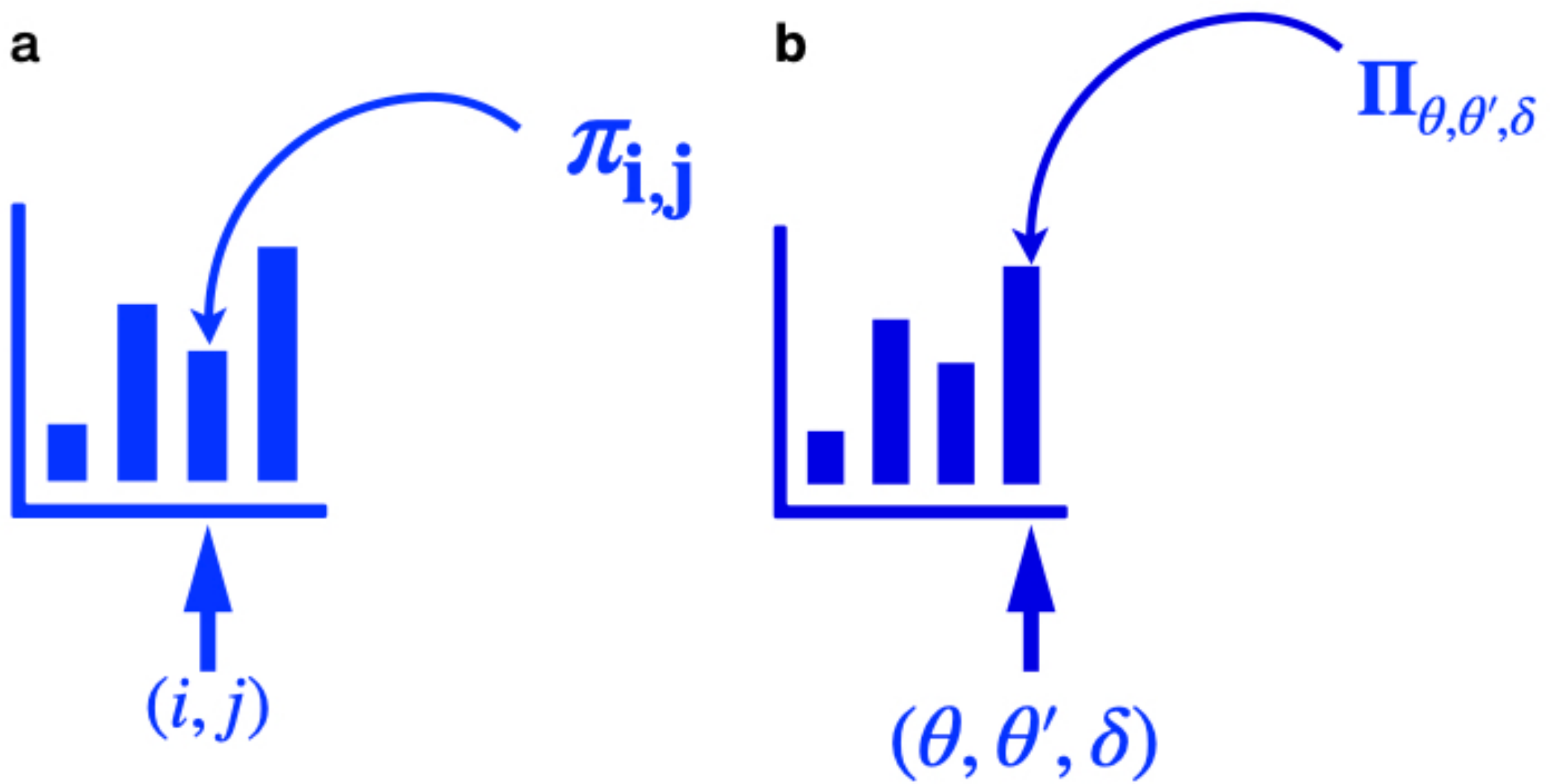}
\caption{ A schematic representation of the classical spatial
    ensembles and their compressed ensemble is shown. (a) The classical spatial ensemble is defined by the probability $\pi_{ij}$ that a link connects node $i$ at one end and node $j$ at the other end.  (b) 
If the latent variable of ecah links are ${\bf X}=(\theta,\theta',\delta)$ the compressed representation of the ensemble in panel (a)  is defined by
the probability $\Pi_{\theta,\theta',\delta}$ that a link connects a node of
 latent variable $\theta$ at one end and a node of latent variable  $\theta'$ at distance ${\delta}$
from the first node at the other end.  Reprinted figure from \cite{radicchi2019classical}.
}
\label{fig:diagram2}
\end{center}
\end{figure}

We consider a  lossy information channel $\hat{Q}$ that for each link $\vec{\ell}=(i,j)$ erases the information about the  identities of the two linked nodes, i.e. erases  the node labels, and only keeps the information about their latent variables. Therefore  the channel $\hat{Q}$ acts on each message $\vec{\ell}$ by performing the transformation
\bea
{\bf X}(\vec{\ell})={\bf X}_{ij}.
\eea
The probability $\Pi_{\bar{\bf X}}$ that the compressed message is given by ${\bf X}(\vec{\ell})=\bar{\bf X}$ is given by 
\bea
\Pi_{\bar{\bf X}}=\sum_{i<j}\pi_{ij}\tilde{\delta}\left({\bf X}(\vec{\ell}),\bar{\bf X}\right)=\pi\left({\bf \bar{X}}\right)N^2P_{\mathcal V}(\bar{\bf X}).
\label{cla:latvar})
\eea
The compressed channel ${\hat{Q}}$ defines a compressed network ensemble described in Figure $\ref{fig:diagram2}$ whose entropy $H$ is  given by 
\bea
H=-L\sum_{\bf X}\Pi_{\bf X}\ln \Pi_{\bf X}=-L\sum_{\bf X}N^2\pi\left({\bf
  X}\right)P_{{\mathcal V}}({\bf X})\ln \left(N^2\pi\left({\bf X}\right) P_{{\mathcal
      V}}({\bf X})\right).
\eea
Interestingly, it can be shown that the entropy $H$ is equal to the mutual information between the input message and the output messages of the channel $\hat{Q}$ multiplied by $L$ (see \cite{radicchi2019classical}).

The {\em optimal latent variables distribution}  $P_{\mathcal V}^{\star}({\bf S})$ is found by considering the following  optimization principle,
\bea
P_{{\mathcal V}}^{\star}({\bf x})=\arg\max_{{P_{\mathcal V}}({\bf x}),\lambda} [H-\lambda S],
\label{principle}
\eea 
under the constraints the  network contains a given number 
of nodes and links,  and that $S=S^{\star}$.
Theresfore the considered optimization principle   consists in maximizing $H$
for a fixed value of  the information content of the network ensemble, i.e. for $S=S^{\star}$.
Since $H$ is proportional to the mutual information of the channel
$\hat{Q}$,
the maximum of $H$ given $S=S^{\star}$ consists in the  capacity of
the channel under the constraint that the network ensemble has entropy $S=S^{\star}$. 
The considered optimization principle for obtaining the  optimal latent variable distribution can be seen as
a parallel of the {\it optimal input distribution}
\cite{mackay2003information} of a channel, with the difference that here we consider a network model with $\pi\left({\bf X})\right)$   fixed 
and we optimize only the distribution of the latent variables ${P_{\mathcal V}}({\bf X})$ given the expected number of links.  
This optimization principle can be also related to the information bottleneck principle in Machine Learning (see Ref. \cite{tishby2000information}).

\subsubsection{Optimal degree distribution}
\begin{figure}
\begin{center}
\includegraphics[width=0.65\columnwidth]{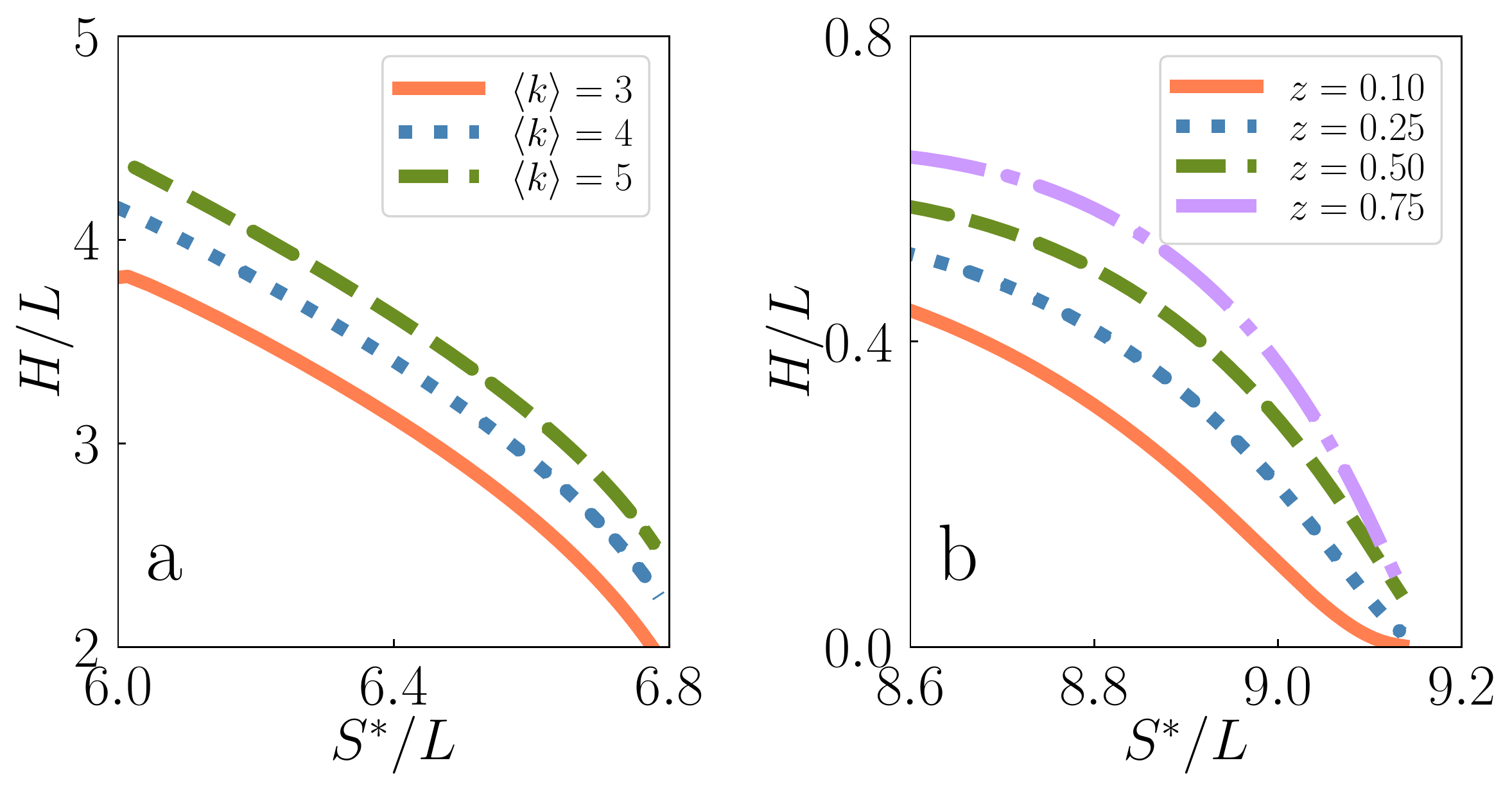}
\caption{ The  entropy $H$ of the compressed network ensemble is plotted as a function of the entropy of the classical network ensemble~$S^{\star}$ for optimal latent variable distribution. In panel (a) we consider the classical network ensembles where only latent variables are the expected degrees of the nodes  and for which the degree distribution $P(k)$ is given by Eq.~(\ref{eq:maximum_entropy_pl}) with different value of the avergae degree $\avg{k}$ of the networks. 
In panel (b) we consider the spatial network ensemble whose only latent variable is the distance $\delta$ between the linked nodes  and the optimal spatial pair correlation function given by Eq.~(\ref{omega:sp}).  In panel (b) we consider  the power-law linking probability  $g(\delta)=\delta^{-\alpha}/z$ with $\alpha=3$ and different  values of~$z$. Reprinted figure from \cite{radicchi2019classical}.
  }
\label{fig:SH}
\end{center}
\end{figure}

In this paragraph we apply the optimization principle defined in the previous paragraph to find the optimal degree distribution.
We consider  the classical network ensemble in which the latent variables are the expected degrees of each network, with expected degree distribution $P(k)$. We recall that this  ensembles has marginals  is given by Eq. (\ref{cla:pikk'}) .
The compressed version of the 
ensemble  is formed by the output of the channel $\hat{Q}$ given by ${\bf X}(\vec{\ell}=(i,j))=(k_i,k_j)$. The probability $\Pi_{\bf X}=\Pi_{k,k'}$ that the lossy compression channel sends a compressed message ${\bf X}(\vec{\ell})=(k,k')$ is given by Eq. (\ref{cla:latvar}) that reads in this case
\bea
\Pi_{k,k'}=\pi(k,k^{\prime})N^2P(k)P(k^{\prime})=\frac{kk'P(k)P(k')}{\langle k\rangle^2}.
\eea
This compressed classical ensemble has entropy $H$  given by
\bea
H=-L\sum_{k,k'}\Pi_{k,k'}\ln \Pi_{k,k'},
\eea
which can be expressed in terms of $P(k)$ as 
\bea
H=-\langle k\rangle N\sum_{k}\frac{kP(k)}{\langle k\rangle}\ln \left(\frac{kP(k)}{\langle k \rangle}\right).
\label{eq:entropy_links}
\eea
In order to find the optimal degree distribution among all the degree distributions with the same explicative power at the node level we maximize
\bea
\max_{P(k),\lambda}\left[H-\lambda \left(S-S^{\star}\right)\right],
\eea
under the constraint that the average degree is constant and the degree distribution is normalised.
To this end, we maximise the functional 
\bea
\mathcal{G}=H-\lambda \left(S-S^{\star}\right)-\xi N\left(\sum_k kP(k)-\langle k \rangle\right)-\rho N\left(\sum_k P(k)-1\right),
\eea
with respect to  $P(k)$ the optimal degree distribution given by 
getting 
\bea
P(k)=Ck^{-(\lambda+1)}e^{-\rho\avg{k}/k}.
\label{eq:maximum_entropy_pl}
\eea
with $C=\avg{k}e^{-(\xi+1)}$ indicating the normalization constant and $\lambda $ and $\rho$ are fixed by the constrain that $S=S^{\star}$ and that the average degree of the network is $\avg{k}$. 
This result is notable as we obtain that the   optimal  degree distribution is scale-free 
with a degree cutoff for small values of the degree providing. Therefore the classical information theory of networks provides an information theory interpretation of the fact that a large fraction of complex networks are scale-free \cite{barabasi1999emergence}.
In figure $\ref{fig:SH}$(a) we plot $H$ versus $S^{\star}$ for the optimal degree distribution $P(k)$ given by Eq.~(\ref{eq:maximum_entropy_pl}).
We observe that the lower is $\lambda$ and therefore the lower is $S^{\star}$ the higher is the entropy $H$. This is because as the power-law exponent  of $P(k)$ decreases  the number of ways to realizes the network decreases but the number of ways to split links  into classes of latent variables  $(k,k')$ increases as the network degrees are more heterogeneous. Therefore this results highlight the information theory benefit of having heterogeneous degree distributions.

\subsubsection{Optimal  distribution of latent variables in classical spatial networks}

In this paragraph we discuss the optimal latent distribution of classical spatial network ensembles.
Let us consider the spatial network ensemble with marginal $\pi_{ij}=\pi(\delta_{ij})$ given by Eq. (\ref{cla:pidelta}). Let us apply to this classical network ensemble the lossy compression channel $\hat{Q}$ transforming any message carrying the information of a link $\vec{\ell}$ in its latent variable ${\bf X}(\vec{\ell}=(i,j))=\delta_{ij}$. The probability $\Pi_{\delta}$ that the channel $\hat{Q}$ transmit the message $\delta$ is given by 
\bea
\Pi_{\delta}=\omega(\delta)g(\delta).
\eea
The output of the channel $\hat{Q}$ can be interpreted as a  compressed spatial network ensemble whose entropy $H$ is given by 
\bea
H=-L\int d{\delta} \Pi_{\delta}\ln \Pi_{\delta}
\label{eq:spatial_H}
\eea

The optimal spatial correlation function $\omega(\delta)$ between nodes can be obtained by applying our optimization principle,  
\bea
\max_{\omega(\delta),\lambda}\left[H-\lambda \left(S-S^{\star}\right)\right],
\eea
under the constraint that  both $\omega(\delta)$ and $\omega(\delta)g(\delta)$ are normalised.
By performing a derivation similar to the one performed in the previous paragraph we obtain that the optimal spatial correlation function is given by 
\bea
\omega(\delta)=C[g(\delta)]^{-\lambda-1}e^{-\rho/g(\delta)},
\label{eq:spatial_pl}
\eea
where $C$ is the normalization constant and where $\lambda$ and $\rho$ impose the constraints $S=S^{\star}$ and $\int d\delta \omega(\delta)g(\delta)=1$.
Interestingly this spatial pair correlation function implies that the optimal spatial distribution of nodes in space is non-uniform.
In the case of power-law function $g(\delta)$ given by
\bea
g(\delta)=\frac{\delta^{-\alpha}}{z}
\eea
the optimal pair correlation function reads
\bea
\omega(\delta)\propto\delta^{\alpha(\lambda+1)}e^{-\rho\delta^{\alpha}/z}.
\label{omega:sp}
\eea
It follows  that the  networks in this ensemble  can be embedded in spatial dimension $D>\alpha (\lambda+1)+1$ where the distribution of points will be fractal with a cutoff at large distances.
As shown in Figure $\ref{fig:SH}$, also for this spatial network ensemble, like for the classical network ensemble whose only latent variables are the expected degrees,   
the entropy $H$ of the compressed ensemble is anti-correlated with the entropy  $S$ when both entropy are evaluated assuming the optimal spatial pair correlation function given by Eq. (\ref{omega:sp}). 

\begin{figure}[!htbp]
\begin{center}
\includegraphics[width=1\textwidth]{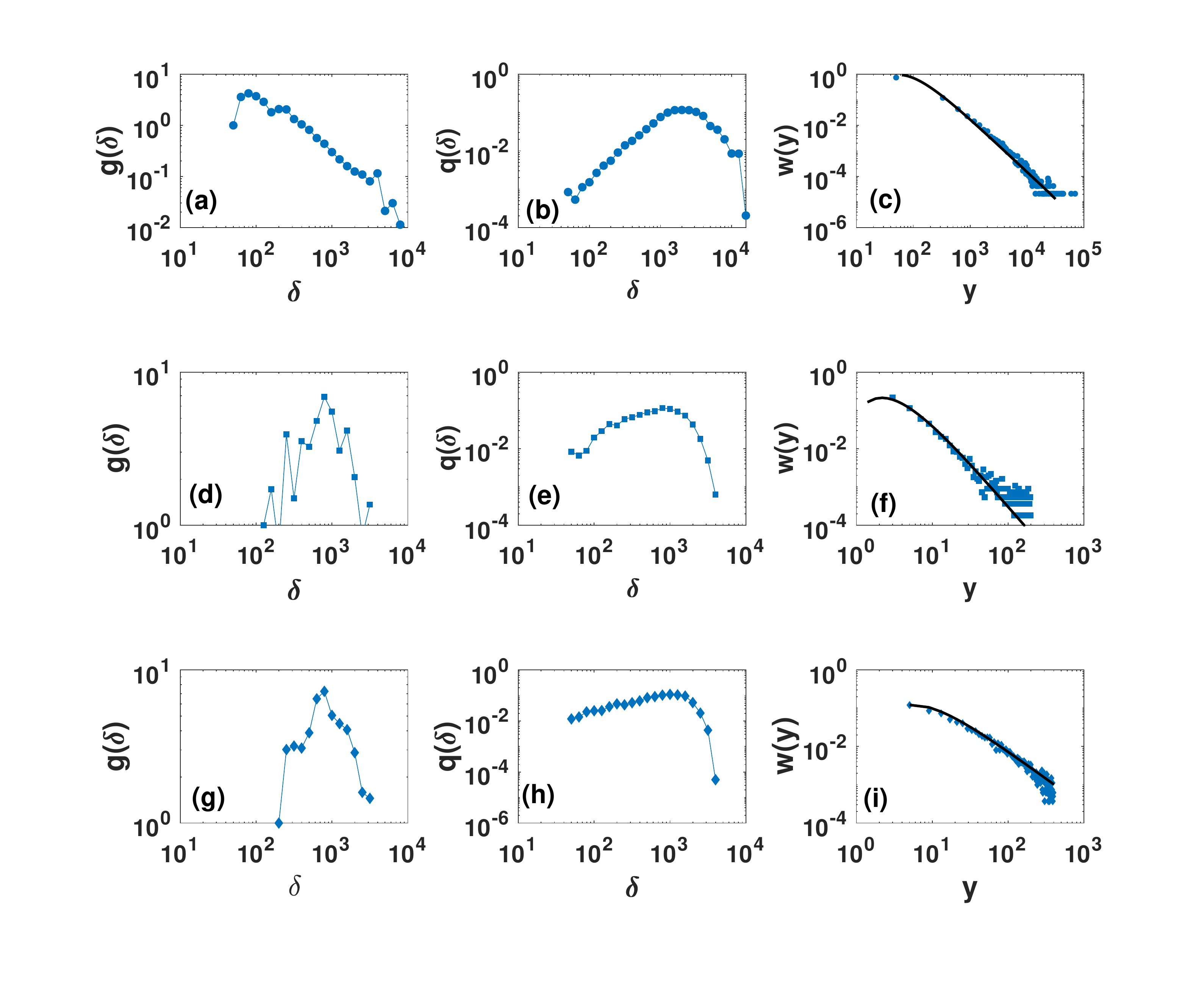}
\caption{ Here we  apply  the classical information-theory framework to real-world airport networks.
The networks correspond to the flights operated by American Airlines (AA)
(panels (a)-(b)-(c))
during January-April 2018 between US airports~\cite{bureau}, by
Lufthansa (LU) (panels (d)-(e)-(f)) and Ryanair (RY) (panels (g)-(h)-(i)) during year 2011 between European airports ~\cite{cardillo}.
For each air carrier,
a separate air transportation network is built, in which nodes are
airports and two airports are connected if at least one flight
between the two airports is present in the data. Using the
network topology and the geographic locations of the airports, 
the empirical linking probability $g(\delta)$ 
and the density distribution $q(\delta)=\int d\theta \int d\theta'
\omega(\theta,\theta',\delta)$ 
 are computed for the three
networks, where distance~$\delta$ is geographic and is measured in kilometers.
Panels~(c)-(f)-(i) shows the pair correlation functions
$\omega(\theta,\theta',\delta)=\omega(y)$, where $y=\theta\theta'g(\delta)$, for the three networks.
Points represent empirical densities, while the full lines are
theoretical predictions according to Eq.~(\ref{eq:th125}). Values of the
Lagrange multipliers are:  $\lambda = 1.2$ and $\nu = 120$ for AA,
$\lambda = 1.3$ and $\nu =5 $ for LU, and $\lambda =0.45$ and $\nu = 8$ for RY. Reprinted figure from \cite{radicchi2019classical}.}
\label{fig:class_airports}
\end{center}
\end{figure}

Finally let us consider  the ensemble in which each link $\vec{\ell}=(i,j)$ has probability $\pi_{ij}$ dictated by the latent variables ${\bf X}=(\theta,\theta',\delta)$ as defined by Eq. (\ref{sp:m:2}).
For each link $\vec{\ell}=(i,j)$ of the original network the lossy compression channel  sends the compressed message ${\bf X}(\vec{\ell})=(\theta_i,\theta_j,\delta_{ij})$. The probability that the channel $\hat{Q}$ sends a message $({\theta,\theta',\delta})$ is given by  
\bea
\Pi_{\theta,\theta',\delta}=\omega(\theta,\theta',\delta)\theta \theta'g(\delta),
\eea
and the entropy of the resulting compressed network ensemble is given by 
\bea
H=-L\int d{\delta} \int d\theta \int d\theta' \Pi_{\theta,\theta',\delta}\ln \Pi_{\theta,\theta',\delta}.
\label{eq:spatial_H2}
\eea
The optimal latent variable distribution  $\omega(\theta,\theta,\delta)$  can be obtained by applying our optimization principle,  
\bea
\max_{\omega(\theta,\theta,\delta),\lambda}\left[H-\lambda \left(S-S^{\star}\right)\right]
\eea
under the constraint that  both $\omega(\theta,\theta,\delta)$ and $\omega(\theta,\theta,\delta)\theta\theta'g(\delta)$ are normalised.
By proceeding similar to the previously discussed cases we find that  $\omega(\theta,\theta,\delta)$ is only a function of $y=\theta\theta'g(\delta)$ and is given by
\bea
\omega(y=\theta\theta'g(\delta))=C\left[\theta\theta'g(\delta)\right]^{-\lambda-1}\exp\left[{-\frac{\nu}{\theta\theta'g(\delta)} }\right].
\label{eq:th125}
\eea
The real data on airport networks indicates that  this optimization principle is satisfied in this class of spatial networks (see Figure \ref{fig:class_airports}).

\section{Conclusions}

In conclusion in this chapter we have discussed the state-of the art of information theory of networks focusing  on spatial networks and in general networks with latent variables.
The field is among the most vibrant of network science as networks are ultimately a way for complex systems to encode  their information.
We have discussed how recently the field of maximum entropy models of networks has been able to significantly evolve thanks to the distinction between microcanonical and canonical ensembles. This distinction is very important to clarify the relation between very well known models of networks such as the configuration model and the exponential random graph. Although this classification is inspired by the theoretical physics field of statistical mechanics, maximum entropy network models display very commonly the non-equivalence of the conjugated ensembles (observed when the number of constraints are extensive) which is instead not the norm in statistical mechanics.
Inspired by statistical mechanic it is also  possible to formulate a classical information theory of networks.
The classical information theory of networks constitutes a very  recent development of the field and  provides for the first time  an information theory explanation  of the origin  of the universal properties of networks. In particular it gives and information theory ground to the wide occurrence of  scale-free degree distribution and  to the  non-uniform distribution in space of the nodes of spatial networks.

The classical information theory of networks is able to predict the optimal latent variable distribution and opens the way for many relevant applications. A  very promising line of research is to explore  the effect of using the optimal latent variable distribution as a prior for inference problems.

\bibliographystyle{unsrt}
\begin{small}
\bibliography{references_spatial}
\end{small}
\end{document}